\documentclass[
reprint,
aps,
prb,
amssymb, amsmath,superscriptaddress,showpacs,
notitlepage,
twocolumn,
footinbib]{revtex4-1}
\usepackage{graphicx} 
\usepackage{pifont}
\usepackage{hyperref}
\usepackage{float}
\usepackage{bm}
\hypersetup{ 
    colorlinks=true, 
    linktoc=all,     
    linkcolor=blue,  
    citecolor =blue
} 
\usepackage[T1]{fontenc}
\usepackage{bm}
\let\vaccent=\v 

\begin{document}
\title{Quantum Criticality of Semi-Dirac Fermions in 2+1 Dimensions}

\author{Mikolaj D. Uryszek}
\affiliation{London Centre for Nanotechnology, University College London, Gordon St., London, WC1H 0AH, United Kingdom}
\author{Elliot Christou}
\affiliation{London Centre for Nanotechnology, University College London, Gordon St., London, WC1H 0AH, United Kingdom}
\author{Akbar Jaefari}
\affiliation{Department of Physical and Biological Sciences, Western New England University, Springfield MA 01119, USA}
\author{Frank Kr\"uger}
\affiliation{London Centre for Nanotechnology, University College London, Gordon St., London, WC1H 0AH, United Kingdom}
\affiliation{ISIS Facility, Rutherford Appleton Laboratory, Chilton, Didcot, Oxfordshire OX11 0QX, United Kingdom}
\author{Bruno Uchoa}
\affiliation{Department of Physics and Astronomy, University of Oklahoma, Norman, OK 73069, USA}

\begin{abstract}
Two-dimensional semi-Dirac fermions are quasiparticles that disperse linearly in one direction and quadratically in the other. 
We investigate instabilities of semi-Dirac fermions towards charge, spin-density wave 
and superconducting orders, driven by short-range interactions. We analyze the critical behavior of the Yukawa theories for the 
different order parameters using Wilson momentum shell RG. We generalize 
to a large number $N_f$ of fermion flavors to achieve analytic control in 2+1 dimensions and calculate critical exponents at one-loop order,
systematically including $1/N_f$ corrections. The latter depend on the specific form of the bosonic infrared
propagator in 2+1 dimensions, which needs to be included to regularize divergencies. The $1/N_f$ corrections are surprisingly small, 
suggesting that the expansion is well controlled in the physical dimension. The order-parameter correlations inherit the 
electronic anisotropy of the semi-Dirac fermions, leading to correlation lengths that diverge along the spatial directions with distinct exponents, 
even at the mean-field level. 
We conjecture that the proximity to the critical point may stabilize novel 
modulated order phases. 
\end{abstract}
\maketitle

\section{Introduction}

\begin{figure}[t!]
 \includegraphics[width=\columnwidth]{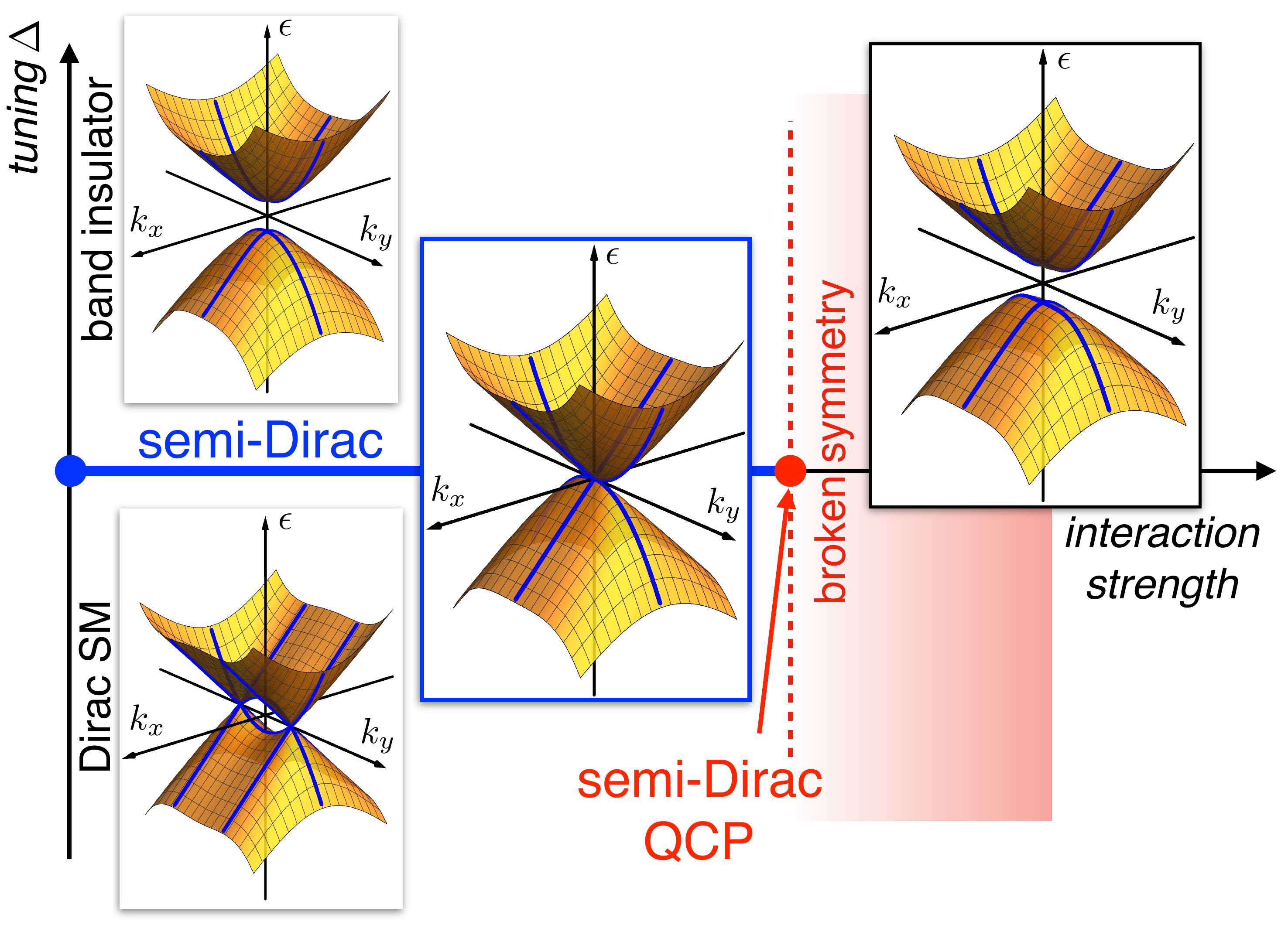}
 \caption{Schematic phase diagram. As a function of the band tuning parameter $\Delta$ 
 the system undergoes a topological  Lifshitz transition between a Dirac semimetal (with a pair of Dirac cones) and a band insulator. At the 
 transition point the system exhibits gapless ``semi-Dirac"  quasiparticle excitations. Sufficiently strong short-range  interactions
 lead to antiferromagnetic, CDW or superconducting order, depending on the type of interaction. The symmetry breaking is associated 
 with the opening of a gap in the semi-Dirac spectrum.}
\label{fig.dispersion}
\end{figure}

Dirac fermions generically describe quasiparticles with relativistic dispersion in the vicinity of special points 
in the Brillouin zone.\cite{CastroNeto2009,Qi2011,Hasan2010,Armitage2018} In two spatial dimensions, 
the merging of two Dirac points results in a topological phase transition separating the semi-metallic
phase from a gapped insulating one.\cite{Dietl+08,Montambaux2009}  At the boundary
between the two phases, the system exhibits gapless ``semi-Dirac''  quasiparticle 
excitations\cite{Banerjee2009} that disperse relativistically in one direction and  quadratically in the other 
(see Fig.~\ref{fig.dispersion}).

Based on density-functional calculations, semi-Dirac quasiparticles were predicted to occur 
in single layers of black phosphorus under strain.\cite{Rodin2014} Shortly after, they were observed
when sprinkling potassium atoms onto single layers of black phosphorus,\cite{Kim2015} 
and more recently using surface doping.\cite{Kim2017} Semi-Dirac electrons have also been 
predicted to occur 
in BEDT-TTF$_{2}$I$_{3}$ salt under pressure,\cite{Katayama2006} VO$_{2}$/TO$_{2}$ 
heterostructures,\cite{Pardo2009,Huang2015} and in strained honeycomb lattices.\cite{Pardo2009}  
Recently, it has been suggested\cite{Christou+18,Christou+19} that second-neighbour repulsions between Dirac 
fermions on the honeycomb lattice can lead to a metallic CDW state that breaks lattice symmetries 
and exhibits semi-Dirac quasiparticles excitations. Semi-Dirac fermions have strongly anisotropic hydrodynamic transport properties,\cite{Link+18}  e.g. the electrical 
conductivity is metallic in one direction and insulating in the other direction. Even more strikingly, one of the electronic sheer viscosity components 
vanishes at zero temperature, leading to a generalization of the previously conjectured lower bound for the viscosity to entropy 
density ratio.\cite{Link+18}

Nodal semi-metals with point-like Fermi surfaces represent the simplest example of fermionic quantum 
criticality,  driven by strong short-range repulsive interactions.\cite{Kotov2012} The symmetry breaking leads to the 
opening of a gap in the fermion spectrum and therefore goes hand-in-hand with a semimetal-to-insulator
transition. In the purely relativistic case of Dirac fermions it is well 
understood\cite{herbutprl2006,herbutetalprb2009,assaadherbutprx2013,Janssen+14} 
that the universal behaviour is described by the Gross-Neveu-Yukawa theory\cite{grossneveuprd1974,zinnjustinnpb1991} 
of chiral symmetry breaking. The coupling between the order-parameter fields and the gapless Dirac fermions 
leads to novel fermion induced critical behaviour that falls outside the Landau-Ginzburg-Wilsom paradigm of a 
pure order-parameter description. Ultimately, the study of quantum phase transitions in nodal semi-metals might serve as a 
stepping stone towards an understanding of quantum criticality in metals with extended Fermi surfaces.

The non-trivial scaling of the quasiparticles kinetic energy of semi-Dirac fermions gives rise to 
novel universal behavior.\cite{Cho2016,Isobe2016,Roy+08,Sur2018} Moreover, it is expected that the inherent electronic anisotropy will be 
reflected in strongly anisotropic order parameter correlations. This can have
profound effects on the nature of broken-symmetry states. For instance, in gapped superconducting phases, it has been suggested\cite{Uchoa2017} that 
an applied magnetic field may lead to the formation a novel smectic phase with a stripe pattern 
of flux domains near the quantum critical point.

Accessing the universal critical behaviour of two-dimensional semi-Dirac fermions has proven difficult. Because of the 
different dispersion along the $k_x$ and $k_y$ directions, the generalization to arbitrary dimension and consecutive 
$\epsilon$ expansion below an upper critical dimension is subtle and not uniquely defined. For a model with $d_\textrm{L}$ 
linear and $d_\textrm{Q}>0$ quadratic momentum directions, the interactions become marginal 
at $2 d_\textrm{L} + d_\textrm{Q} = 4$, suggesting that the universal critical behaviour of 
two-dimensional semi-Dirac fermions ($d_\textrm{L} = d_\textrm{Q} = 1)$ could be accessible 
within an $\epsilon$-expansion with $d_\textrm{L} =1$ and $d_\textrm{Q} = 2-\epsilon$.\cite{Sur2018}
This expansion results in a non-analytic $\sim\epsilon \ln\epsilon$ dependence of the anomalous 
dimensions of the fermion and order-parameter fields.\cite{Sur2018} The non-monotonic
behaviour and vanishing of the corrections at $\epsilon=1$, nevertheless, could indicate that the semi-Dirac case 
lies outside  the validity of the expansion.
 
In a complementary approach\cite{Roy+08} the problem was analyzed in two spatial dimensions with a 
 generalized dispersion $k_y^{2n}$ in the non-relativistic direction. This construction allows for a controlled ascent
 from one dimension ($n\to\infty$). At finite $n$, interactions are rendered irrelevant at weak coupling, but key aspects 
 of one-dimensional physics such as spin-charge separation are preserved.
 Quantum fluctuations beyond 1D, controlled by $\sim1/n$, enter the renormalizatio group (RG) through loop integrations that involve 
 the dispersion along $k_y$. 
 
 In this paper we revisit the criticality of semi-Dirac fermions in 2+1 dimensions.
 We avoid tuning the dimensionality or form of the dispersion but instead introduce a large number 
 $N_f$ of fermion flavors for analytical control. The additional fermion flavors 
are not involved in the symmetry breaking and remain degenerate across the quantum phase transition. 
We use a one-loop, momentum-shell RG to calculate critical 
 exponents to order $1/N_f$. A similar procedure was used to analyze the criticality of 2D and 3D 
 semi-Dirac fermions subject to unscreened, long-range Coulomb interactions.\cite{Yang+14,Isobe2016} Here we focus on 
 short-range interactions that drive antiferromagnetic, CDW and superconducting instabilities. The results can be readily  
 compared to the analogous large $N_f$ analysis of relativistic 2D Dirac fermions,\cite{herbutprl2006,Janssen+14} 
 unravelling the effects of the peculiar form of the dispersion on the universal behavior.
 
We find that the $1/N_f$ corrections to critical exponents are very small and considerably smaller than 
for the case of 2D Dirac fermions, suggesting that the expansion is well controlled. In the mean-field limit, 
$N_f\to\infty$, we recover the results obtained from the $d_Q=2-\epsilon$ expansion,\cite{Sur2018} 
evaluated at $\epsilon=1$ and $N_f\to\infty$. The $1/N_f$ corrections differ however
since they depend on the specific form of the bosonic infrared (IR) propagator in 2+1 dimensions, which needs to be 
included to regularise divergencies. As expected, we find that the order-parameter correlations inherit 
the intrinsic anisotropy of the system, e.g. the correlation lengths along different spatial directions diverge 
with different powers. We conjecture that this behavior could help stabilize exotic modulated ordered phases near
the quantum critical point.

The  $N_f\to\infty$  results are significantly different from the 2D Dirac case. 
This can be understood by analyzing the mean-field Ginzburg-Landau free energy that is obtained from integrating
out the fermions in the broken-symmetry state.  
Spatial anisotropies are encoded in non-analytic gradient terms. 
We find that the critical exponents derived from the RG for $N_f\to\infty$ are in agreement with ones obtained from the mean-field Ginzburg-Landau functional,
suggesting that hyperscaling relations are satisfied. 

The outline of the paper is a follows: in Sec.~\ref{sec.Yukawa} we derive the effective Yukawa actions 
for spin, charge and superconducting instabilities. In Sec.~\ref{sec.RG} we explain the Wilson RG  procedure, 
derive the one loop RG equations  in the large $N_f$ limit and compute critical exponents to order $1/N_f$. The 
non-analytic structure of the mean-field theory is discussed in Sec.~\ref{sec.MF}. Finally, in Sec.~\ref{sec.dis} we 
summarise our results and discuss their implications.

\section{Effective Field Theory}
\label{sec.Yukawa}

In this Section we motivate the effective low-energy field theory for different instabilities of semi-Dirac fermions in 
2+1 dimensions. The part of the action describing non-interacting fermions is given by
\begin{eqnarray}
\label{eqn:s_psi}
\mathcal{S}_{\bm{\psi}} & = & \sum\limits _{n=1}^{N_{f}}\int_{\vec{k}}\bar{\bm{\psi}}_n (\vec{k})  \bm{s}^{0}\otimes\bigg[-ik_{0}\bm{\sigma}^{0}
+vk_{x}\bm{\sigma}^{x}\nonumber\\
& & \hspace{2.5cm}+\bigg(\frac{k_{y}^{2}}{2m}+\Delta\bigg)\bm{\sigma}^{y}\bigg]\bm{\psi}_n(\vec{k}),
\end{eqnarray}
where $\vec{k}=(k_{0},\bm{k})=(k_{0},k_{x},k_{y})$, with $k_{0}$ the Matsubara frequency,
and $\int_{\vec{k}}\equiv\int\frac{d^{3}\vec{k}}{(2\pi)^{3}}$, subject to an ultraviolet cut-off $\Lambda$. 
The two sets of Pauli matrices $\bm{s}^i$ and $\bm{\sigma}^i$ ($i=0$ for identities) 
act on the spin and sublattice spaces, respectively, and the fermionic Grassmann fields 
$\bm{\psi}_n(\vec{k})$ are 4-component spinors. We have generalised the action by introducing 
$N_f$ flavours or copies of Grassmann fields, labelled by $n$.

From the poles of the corresponding Green function we obtain the electron dispersion,
\begin{equation}
\label{eqn:disp}
\epsilon(\bm{k}) = \pm \sqrt{\left(v k_x\right)^2+\left(\frac{k_y^2}{2 m}+\Delta\right)^2},
\end{equation}
which is degenerate in spin $s=\uparrow,\downarrow$ and fermion flavour $n$. The effect of the 
tuning parameter $\Delta$ on the electron dispersion is illustrated in Fig.~\ref{fig.dispersion}. For 
$\Delta < 0$ the dispersion contains two relativistic Dirac points $\bm{K}_\pm=\left(0,\pm\sqrt{2m(-\Delta)}\right)$, 
while for $\Delta>0$ the dispersion has an energy gap $\Delta$. Hence $\Delta$ tunes 
a transition between a Dirac semimetal and a band insulator. At $\Delta=0$, the system undergoes 
a topological Lifshitz transition, corresponding to the merging of two Dirac points. At this point 
the system exhibits semi-Dirac quasiparticle excitations. 

We aim to describe interaction-driven quantum phase transition of semi-Dirac fermions. 
Enforcing $\Delta=0$ while increasing the strength of interactions requires fine tuning. Depending on 
the experimental system, this may be achieved with strain, pressure, or surface 
doping.\cite{Rodin2014,Kim2015,Kim2017,Katayama2006,Pardo2009,Huang2015}

\subsection{CDW and SDW Instabilities}

To study the criticality of semi-Dirac fermions subject to local interactions we use the Yukawa formalism. 
This amounts to performing a Hubbard-Stratonovich transformation of a generic four fermion interaction vertex 
in the appropriate order parameter channel. Here we don't address the question of phase competition 
but instead focus on a particular type of symmetry breaking, and assume an underlying fermion interaction
that would stabilize this order. 

In the case of CDW and SDW instabilities, the Yukawa coupling between the order parameter fields $\phi_i(\vec{k})$
and the fermions $\bm{\psi}_n(\vec{k})$ is given by
\begin{equation}
\label{eqn:yukawa}
\mathcal{S}_{g}=g\sum\limits _{n=1}^{N_{f}}\sum_{i}^{N_{b}}\int_{\vec{k},\vec{q}}  \phi_i (\vec{q}) \, \bar{\bm{\psi}}_n (\vec{k}) \mathbf{Y}_{i}\bm{\psi}_n(\vec{k}+\vec{q}),
\end{equation}
where 
\begin{equation}
\label{eqn:upsilon0}
\mathbf{Y}_{i}=\begin{cases}
\bm{s}^{0}\otimes\bm{\sigma}^{z} & \quad\text{CDW}:i=0\\
\bm{s}^{i}\otimes\bm{\sigma}^{z} & \quad\text{SDW}:i=\{x,y,z\}.
\end{cases}
\end{equation}
In spite of their name, the CDW and the SDW states describe a staggered
field in the pseudospin space ($\sigma^{z}$), which could in principle
be any generic quantum number in the original lattice model, such
as sublattice, valley or orbital quantum numbers. In the CDW state,
the order parameter is a scalar ($N_{b}=1$), whereas in the SDW,
which in this language is equivalent to antiferromagnetism, $N_{b}=3$.

In addition to a $\bm{\phi}^2$ term that arises from the Hubbard-Stratonovich 
transformation, the successive elimination of high-energy fermion modes
under the renormalization group will generate gradient terms, as well as a 
$\bm{\phi}^4$ vertex.  The resulting 
Ginzburg-Landau functional for the order parameter is given by 
$\mathcal{S}_{\bm{\phi}}+\mathcal{S}_\lambda$ with

\begin{eqnarray}\label{eqn:s_phi}
\mathcal{S}_{\bm{\phi}} & = & \frac{1}{2}\int_{\vec{q}}\left( c_{0}^{2}q_{0}^{2}+c_{x}^{2}q_{x}^{2}+c_{y}^{2}q_{y}^{2}+m_{\phi}^{2}\right) \big|\bm{\phi}(\vec{q})\big|^{2},\\
\mathcal{S}_\lambda & = & \lambda \int_{\bm{r},\tau} |\bm{\phi}(\bm{r},\tau)|^4,
\end{eqnarray}
where the integral of the $\bm{\phi}^4$ vertex runs over real space $\bm{r}=(x,y)$ and imaginary time $\tau$. The order parameter mass term is the tuning parameter for the broken symmetry state: at the critical 
point $m_{\phi}^{2}=0$.
To summarize, the effective field theory describing the criticality of semi-Dirac 
fermions in 2+1 dimensions is given by the sum of four terms, 
\begin{equation}\label{eqn:full_action}
\mathcal{S}=\mathcal{S}_{\bm{\psi}}+\mathcal{S}_{\bm{\phi}}+\mathcal{S}_{g}+\mathcal{S}_{\lambda}.
\end{equation}

There is a caveat, however.  As we will discuss in Sec.~\ref{sec.RG}, the bosonic propagator $G_{\bm{\phi}}(\vec{q})$ 
develops an unphysical singularity under the renormalization group scheme. It is therefore necessary to 
regularize this divergence by including an IR contribution in $\mathcal{S}_{\bm{\phi}}$.

\subsection{Superconducting Instability}

We also analyze a possible instability in the superconducting channel.
In that case, we initially consider a generic Hamiltonian 
$\mathcal{H}_{0}(\bm{p})$ with semi-Dirac quasiparticles that preserves time 
reversal symmetry (TRS), such that $\mathcal{T}\mathcal{H}_{0}(\bm{p})\mathcal{T}^{-1}=\mathcal{H}_{0}(\bm{p})$,
with $\bm{p}$ the momentum away from the center of the zone.
We restrict our attention to families of TRS Hamiltonians that support an isotropic
pairing gap around the semi-Dirac points. That is allowed when pairing
occurs across two semi-Dirac points sitting in opposite sides of
the Brillouin zone, such that each Cooper pair has zero total momentum.
A concrete example of a TRS tight-binding Hamiltonian with a pair of semi-Dirac points can be 
found in Ref.~\onlinecite{Uchoa2017}.

Due to TRS, the generic Hamiltonian can be written in a Bogoliubov
de Gennes basis $\left(c_{\bm{p},s},c^\dagger_{-\bm{p},-s}\right)$ 
with $s$ the spin index as
\begin{equation}\label{eqn:tau2}
\mathcal{H}_{\text{BdG}}(\bm{p})=\mathcal{H}_{0}(\bm{p})\otimes\tau^{z},
\end{equation}
where $\tau^{z}$ is a Pauli matrix in the Nambu space. Expanding this Hamiltonian 
around one of the semi-Dirac points described, the action in the Nambu basis is
\begin{eqnarray}
\label{eqn:Spsi2}
\mathcal{S}_{\bm{\psi}} & = & \sum\limits _{n=1}^{N_{f}}\int_{\vec{k}}\bar{\bm{\psi}}_n (\vec{k})  \bm{s}^{0}\otimes\bigg[-ik_{0}\bm{\sigma}^{0}\otimes\bm{\tau}^0
+vk_{x}\bm{\sigma}^{x}\otimes\bm{\tau}^z\nonumber\\
& & \hspace{2.cm}+\bigg(\frac{k_{y}^{2}}{2m}+\Delta\bigg)\bm{\sigma}^{y}\otimes\bm{\tau}^z\bigg]\bm{\psi}_n(\vec{k}),
\end{eqnarray}
where we enforce $\Delta=0$ at the fixed point. The expansion around
the opposite Dirac point gives an equivalent copy of the action above, which  
can be accounted for in the fermionic degeneracy $N_{f}$. 

The Yukawa coupling between the complex, two-component ($N_{b}=2$) order parameter 
of the $s$-wave superconductor and the fermions is given by Eq.~(\ref{eqn:yukawa}) with 
the coupling matrix
\begin{equation}\label{eqn:Upsilon}
\mathbf{Y}_{i}=\bm{s}^{0}\otimes\bm{\sigma}^{0}\otimes\bm{\tau}^{i}\quad\text{SC}:i=\{x,y\}
\end{equation}
in the enlarged Nambu space. $\mathcal{S}_{\bm{\phi}}$ and $\mathcal{S}_\lambda$ take the same 
form as in the CDW/SDW case.

\section{Renormalization Group Analysis}
\label{sec.RG}

In this Section, we outline the RG approach used to analyze the universal critical behaviour of the Yukawa actions derived in Sec.~\ref{sec.Yukawa} for 
CDW, SDW and superconducting instabilities. In the case of the superconducting instability, one is required to include the Nambu space, as indicated
in the action (\ref{eqn:Spsi2}), and sum over half the number of states in the trace to avoid double counting. For simplicity,  we omit any explicit 
mention to the Nambu space, which can be trivially incorporated for the SC instability.

The universal critical behaviour, e.g. the critical exponents,  should not depend on the cut-off scheme. In the following we treat frequency and momentum 
on an equal footing and impose the ultraviolet (UV) cut-off $\Lambda$ in 2+1 dimensional momentum-frequency space,
\begin{equation}
\int_{\vec{k}}\equiv \int \frac{d^{3}\vec{k}}{(2\pi)^{3}}\ \theta\left(\Lambda- \sqrt{k_0^2+\epsilon^2(\bm{k})}\right),
\end{equation}
where the electron dispersion $\epsilon(\bm{k})$ is defined in Eq.~(\ref{eqn:disp}) with $\Delta=0$. We will analyze the RG flow of the coupling 
constants when successively integrating out high-energy modes with momenta and frequencies from the infinitesimal shell 
\begin{equation}
\label{eqn:shell}
\Lambda e^{-z d\ell} < \sqrt{k_0^2+\epsilon^2(\bm{k})}< \Lambda,
\end{equation}
followed by a rescaling of frequency and momenta at each step, 
\begin{equation}
\label{eqn:rescaling}
k_{0}=k_{0}^{\prime}e^{-z\,d\ell},\; k_{x}=k_{x}^{\prime}e^{-z\,d\ell}, \;k_{y}=k_{y}^{\prime}
e^{-z_{y}\,d\ell}.
\end{equation}

In the definitions of the 2+1 dimensional shell (\ref{eqn:shell}) and the rescaling (\ref{eqn:rescaling}), 
frequency $k_0$ and the momentum $k_x$ along the relativistic direction 
are treated on an equal footing and are both rescaled with a dynamical 
exponent $z$ relative to the $k_y$ direction. One might view this as having one space-like and 
two time-like directions. For greater clarity, we have introduced a scaling dimension $z_y$ 
of the momentum $k_y$, which we will set to one later. Under successive mode decimation 
and rescaling, the cut-off remains invariant if 
\begin{equation}
z=2z_{y}.
\end{equation}
The shell integration is performed using the coordinate transformation 
\begin{eqnarray}
\label{eqn:para}
k_{0} & = & \epsilon\sin\theta\cos\phi,\nonumber\\
vk_{x} & = & \epsilon\sin\theta\sin\phi,\\
\frac{k_{y}^{2}}{2m} & = & \epsilon\cos\theta,\nonumber
\end{eqnarray}
with $\epsilon\in[\Lambda e^{-z d\ell}  ,\Lambda]$, $\theta\in[0,\frac{\pi}{2}]$, and
$\phi\in[0,2\pi]$. The Jacobian of the transformation is
\begin{equation}\label{eqn:rho2}
\rho(\epsilon,\theta,\phi)=\frac{\sqrt{2m}}{2v}\frac{\sin\theta}{\sqrt{\cos\theta}} \epsilon^{3/2}.
\end{equation}

\subsection{Tree Level Scaling}
\label{sec.scaling}

Let us first consider the consequences of the rescaling of frequency and momenta (\ref{eqn:rescaling}) 
on the free fermion action $\mathcal{S}_{\bm{\psi}}$ (\ref{eqn:s_psi}). 
Since frequency $k_0$ and momentum $k_x$ have the same scaling dimensions, $[k_0]=[k_x]=z$, 
the velocity $v$ of the relativistic dispersion along $k_x$ does not flow. The relation $z=2 z_y$ ensures
that the mass $m$ associated with the quadratic $k_y$ dispersion remains constant at the tree level.

In order to keep the overall prefactor of $\mathcal{S}_{\bm{\psi}}$ constant, we rescale the fermion fields 
as 
\begin{equation}
\bm{\psi}_n(\vec{k})=\bm{\psi}_n^{\prime}(\vec{k^{\prime}})\ e^{d\ell(3z+z_{y}-\eta_{\psi})/2}.
\end{equation}
Here $\eta_{\psi}$ is the anomalous dimension of the fermion fields, which will be used to absorb the 
renormalization due to the decimation of high energy modes.

Let us turn our attention to the quadratic bosonic action $\mathcal{S}_{\bm{\phi}}$ (\ref{eqn:s_phi}).
The distinct scaling dimensions of momenta imply that the coefficients $c_0^2$ and $c_x^2$ will scale 
differently from $c_y^2$. If we rescale the bosonic fields as 
\begin{equation}
\bm{\phi}(\vec{q})=\bm{\phi}^{\prime}(\vec{q^{\prime}})e^{d\ell(2z+3z_{y}-\eta_{\phi})/2},
\end{equation}
where $\eta_{\phi}$ is the anomalous dimension of the $\phi$ field, the $c_y^2$ coefficient does
not flow at tree level. Under this rescaling both $c_{0}^{2}$ and $c_{x}^{2}$
are irrelevant at tree level with scaling dimensions $[c_{0}^{2}]=[c_{x}^{2}]=-z$.
Both parameters flow to zero and can be omitted from the bare bosonic propagator. This omission 
poses problems, as the propagator no longer depends on $q_{0}$ and $q_{x}$, leading to unphysical 
divergencies in the infrared (IR) limit. That issue however can be
fixed with an IR regularization, as discussed in the next Subsection (\ref{sec:IR}).

The above field rescaling leads to the following tree level scaling
dimensions for the Yukawa and the $\phi^{4}$ couplings,
\begin{equation}\label{eqn:g}
[g]=\frac{z_y}{2},\qquad[\lambda]=-z_y.
\end{equation}
This portrays that the $\bm{\phi}^{4}$ term is irrelevant and can be
discarded, while the Yukawa coupling is a relevant perturbation at
tree level.

\subsection{Infrared Regularization}
\label{sec:IR}

In order to regularize unphysical divergencies of the bosonic propagator in the limit $q_{y}\to0$, we augment it by 
an IR contribution.\cite{Li2018,Han2018,Han2019}  As pointed out before,\citep{Isobe2016} in 2+1 dimensions the bosonic propagator has different 
asymptotic forms in the UV and IR limits. 

\begin{figure}[t]
 \includegraphics[width=0.4\columnwidth]{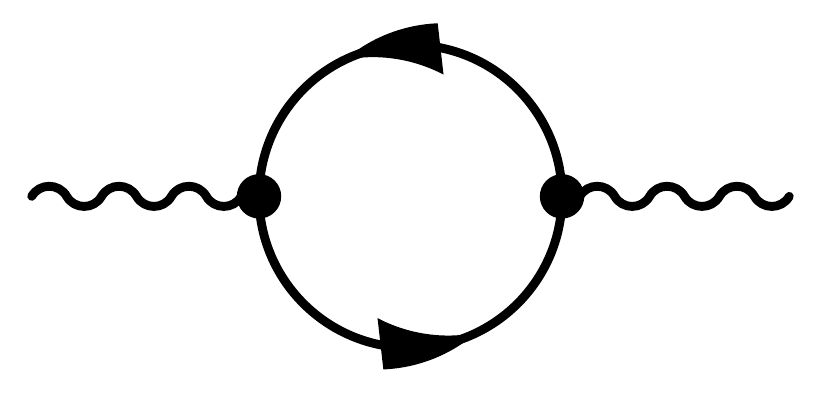}
 \caption{Polarization bubble diagram describing the IR regulator of
the bosonic propagator and the self-energy correction of the bosons
in the momentum shell.}
\label{fig:b_se}
\end{figure}

Since the RG flow is generated by a successive integration of modes from a shell near the UV cut-off, the IR contribution
is not generated or renormalized under the RG. Instead it needs to be computed separately by integrating the 
fermion polarization (see Fig.~\ref{fig:b_se}) over the \emph{full} frequency and momentum range,\cite{Isobe2016} 
\begin{equation}\label{eqn:IR}
\Pi_{ij}(\vec{q})=\frac{g^{2}}{2}\text{Tr}\int_{\vec{k}}\mathbf{Y}_{i}\mathbf{G}_{\psi}(\vec{k})\mathbf{Y}_{j}\mathbf{G}_{\psi}(\vec{k}+\vec{q}),
\end{equation}
where $\mathbf{Y}_{i}$ is the Yukawa coupling for a given instability, as defined in Eqs. (\ref{eqn:upsilon0}) and 
(\ref{eqn:Upsilon}), and 
\begin{equation}
\label{eqn:GPsi0}
\mathbf{G}_{\psi}(\vec{k})=\bm{s}^{0}\otimes\frac{ik_{0}\bm{\sigma}^{0}+vk_{x}\bm{\sigma}^{x}+\frac{k_{y}^{2}}{2m}\bm{\sigma}^{y}}{k_{0}^{2}+(vk_{x})^{2}+k_{y}^{4}/(2m)^{2}}
\end{equation}
is the fermionic propagator.  

For the multicomponent order parameters of the SDW and superconducting phases, the polarization 
is diagonal, $\Pi_{ij}(\vec{q})=\Pi(\vec{q})\delta_{ij}$, reflecting the underlying O(3) and U(1) symmetries. 
In the limit of small $q_{0}$, $q_{x}$ and at $q_{y}=0$, the leading term is
\begin{equation}
\label{eqn:IR_prop}
\Pi_{\text{IR}}(\vec{q})=N_{f}g^{2}\frac{\sqrt{2m}}{v}\left(q_{0}^{2}+v^{2}q_{x}^{2}\right)^{\frac{1}{4}},
\end{equation}
where $N_{f}$ is the number of fermion flavors. This contribution to the kernel of $\mathcal{S}_{\bm{\phi}}$
regularizes the bosonic propagator in the IR at the critical surface $(m_\phi^2=0)$, 
\begin{equation}
\label{eqn:bosonic_prop}
G_{\phi}^{-1}(\vec{q})=N_{f}g^{2}\frac{\sqrt{2m}}{v}\bigg(q_{0}^{2}+v^{2}q_{x}^{2}\bigg)^{\frac{1}{4}}+c_{y}^{2} q_{y}^{2}.
\end{equation}

\subsection{Self-Energy and Vertex Corrections}

Using the propagators (\ref{eqn:GPsi0}) and  (\ref{eqn:bosonic_prop}) for fermionic  and bosonic  fields, respectively, 
we can now go beyond the tree level scaling and extract one-loop corrections to the propagators and the 
Yukawa coupling. As mentioned before, the bosonic $\bm{\phi}^4$ is irrelevant and can be dropped. 

We first concern ourselves with the one-loop renormalization of the regularized bosonic 
propagator (\ref{eqn:bosonic_prop}). The only component that is of interest is in the $q_{y}$ direction
as the dependence on linear momentum and frequency directions in the propagator comes from the IR, 
which is not renormalised under the RG.  The one-loop bosonic self-energy is depicted in Fig.~\ref{fig:b_se}
and takes the form

\begin{equation}
\label{eqn:PI>}
\Pi^{>}(\vec{q}) = \frac{g^{2}}{2}\text{Tr}\int_{\vec{k}}^{>}\mathbf{Y}_{i}\mathbf{G}_{\psi}(\vec{k})\mathbf{Y}_{i}\mathbf{G}_{\psi}(\vec{k}+\vec{q}),
\end{equation}
where $\int_{\vec{k}}^{>}$ means integration over the UV modes within the frequency-momentum shell of width $\Lambda z\,d\ell$,
as defined in Eq.~(\ref{eqn:shell}). Note that in the above expression, the order-parameter field component $i$ is not 
summed over and that the result is the same for all components.  The leading terms of the self-energy have the form 
$\Pi^{>}(\vec{q})=\Pi_{0}q_{0}^{2}+\Pi_{x}q_{x}^{2}+\Pi_{y}q_{y}^{2}+\Pi_{m_{\phi}^{2}}$.
Expanding the fermionic propagator to second order in $q_y$, then performing
the coordinate transformation (\ref{eqn:para}) and integrating over the energy shell, we find that the renormalization of the $c_y^2$ coefficient 
is given by 
\begin{equation}
\label{eqn:deltacy}
d(c_{y}^{2})=2N_{f}\frac{11}{21\pi^{2}}\frac{\sqrt{2m}}{v}\frac{g^{2}}{\sqrt{\Lambda}}z d\ell\equiv\Pi_{y}z d\ell.
\end{equation}
Although the mass of the bosons $m_\phi^2$ also runs in the RG flow, for now it will be fined tuned to zero at the critical surface. We will address the renormalization of $m_\phi^2$ in detail later on in sec.  \ref{sec:critical_exp} and \ref{sec:multicriticality}, when we examine the vicinity of the quantum multicritical point.

\begin{figure}[t]
 \includegraphics[width=0.8\columnwidth]{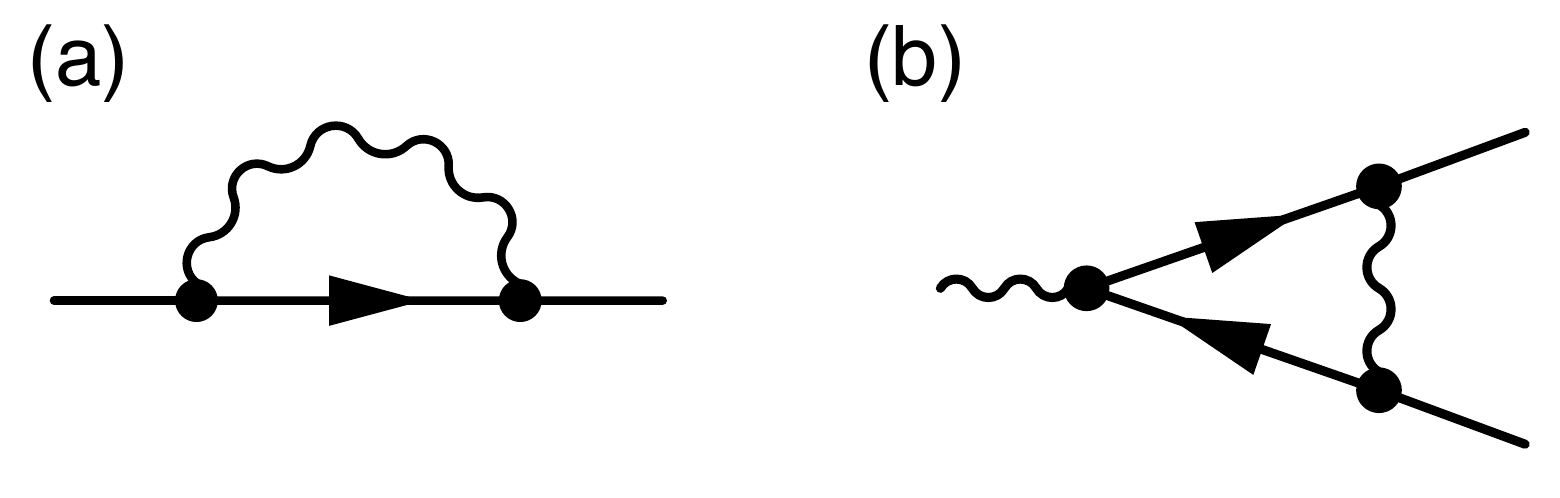}
 \caption{(a) Self-energy correction to the fermionic propagator in one-loop. (b) Vertex correction diagram to the Yukawa coupling.
 The bosonic propagator is represented by the wavy line while the fermionic propagator by the straight line.} 
\label{fig:vertex}
\end{figure}

Next, we turn our attention to the one-loop fermionic self energy [Fig. \ref{fig:vertex}(a)], which is 
equal to 
\begin{equation}\label{eqn:sigma2}
\bm{\Sigma}(\vec{k})=-g^{2}\sum\limits _{i}^{N_{b}}\int_{\vec{q}}^{>}G_{\phi}(\vec{q})\mathbf{Y}_{i}\mathbf{G}_{\psi}(\vec{k}+\vec{q})\mathbf{Y}_{i}.
\end{equation}
After shell integration, it takes the form 
\begin{equation}
\label{eqn:SIgmamatrix}
\bm{\Sigma}(\vec{k})=\bm{s}^{0}\otimes[\Sigma_{0}(k_{0}\bm{\sigma}^{0}+vk_{x}\bm{\sigma}^{x})+\Sigma_{y}\frac{k_{y}^{2}}{2m}\bm{\sigma}^{y}]z d\ell,
\end{equation}
where
\begin{align}
\Sigma_{0}& =\frac{N_{b}}{2N_{f}}F_{1}\bigg(\frac{2N_{f}\sqrt{2m}g^{2}}{vc_{y}^{2}\sqrt{\Lambda}}\bigg)  \label{eqn:Sigma0}\\
\Sigma_{y} & =\frac{N_{b}}{2N_{f}}F_{2}\bigg(\frac{2N_{f}\sqrt{2m}g^{2}}{vc_{y}^{2}\sqrt{\Lambda}}\bigg),\label{eqn:Sigmay}
\end{align}
with $\Sigma_x=\Sigma_0$, and 
\begin{align}
F_{1}(x) & =\frac{1}{4\pi^{2}}\int_{0}^{\frac{\pi}{2}}d\theta\ \frac{(\cos\theta)^{\frac{3}{2}}\sin\theta}{x^{-1}\cos\theta+\sqrt{\sin\theta}},\label{eqn:F1}\\
F_{2}(x) & =\frac{1}{4\pi^{2}}\int_{0}^{\frac{\pi}{2}}d\theta\ \frac{\cos2\theta+2\cos4\theta}{x^{-1}\cos\theta+\sqrt{\sin\theta}}\frac{\sin\theta}{\sqrt{\cos\theta}}\label{eqn:F2}
\end{align}
are defined as integral functions. 

The renormalization of the Yukawa vertex at one-loop order is obtained from the diagram shown in Fig. \ref{fig:vertex}(b) with the 
external frequencies and momenta set to zero,
\begin{equation}
\label{eqn:Gamma}
\bm{\Gamma}_{i}=g^{3}\sum_{j}^{N_{b}}\int_{\vec{q}}^{>}G_{\phi}(\vec{q})\mathbf{Y}_{j}\mathbf{G}_{\psi}(\vec{q})\mathbf{Y}_{i}\mathbf{G}_{\psi}(\vec{q})\mathbf{Y}_{j}.
\end{equation}
The matrix $\bm{\Gamma}_{i}$ is proportional to the Yukawa matrix $\mathbf{Y}_i$, $\bm{\Gamma}_{i} = g\, \Omega \,\mathbf{Y}_{i}z d\ell$, where we have absorbed a factor 
of $g^2$ in the definition of $\Omega$. Performing 
the shell integral we obtain 
\begin{equation}
\label{eqn:Omega}
\Omega = -\frac{2-N_{b}}{2N_{f}}F_{3}\left(\frac{2N_{f}\sqrt{2m}g^{2}}{vc_{y}^{2}\sqrt{\Lambda}}\right)
\end{equation}
with
\begin{equation}\label{eqn:F3}
F_{3}(x)=\frac{1}{4\pi^{2}}\int_{0}^{\frac{\pi}{2}}d\theta\ \frac{1}{x^{-1}\cos\theta+\sqrt{\sin\theta}}\frac{\sin\theta}{\sqrt{\cos\theta}}.
\end{equation}
Note that the vertex correction to the Yukawa coupling $g$ has opposite sign for the CDW ($N_b=1$) and SDW ($N_b=3$) 
instabilities and vanishes in the case of a superconductor ($N_b=2$), as reported in previous studies.\cite{Sur2018}
Due to the linear dispersion in the $k_x$ direction, the vertex correction does not generate new couplings in the RG flow, as 
 it happens  in nodal metals with purely quadratic touching points.\cite{Vafek2010}

\subsection{RG Equations}
\label{sec:implicit_RG} 

As a first step we analyze the RG flow of the fermion propagator $\mathcal{S}_{\bm{\psi}}$ (\ref{eqn:s_psi}), 
combing the contributions from the self-energy loop integral $\bm{\Sigma}$ (\ref{eqn:sigma2}) and the rescaling 
outlined in Sec.~\ref{sec.scaling}. Since $\Sigma_0=\Sigma_x$, the coefficients of the $k_0$ and $k_x$ terms are 
renormalized in same way. We can keep the prefactor of the two terms constant if we define the anomalous dimension 
of the fermion fields as 
\begin{equation}\label{eqn:eta_psi}
\eta_{\psi}= z\Sigma_{0}.
\end{equation}
The resulting RG equations for $m^{-1}$ is then given by
\begin{equation}
\frac{d(m^{-1})}{d\ell}  =  m^{-1}\left(z-2z_{y}+z \Sigma_{y}-z \Sigma_{0}\right),
\end{equation}
with the self-energy corrections $\Sigma_{0}$ and $\Sigma_{y}$ given in Eqs.~(\ref{eqn:Sigma0}) and (\ref{eqn:Sigmay}), respectively. 
In order to keep the full fermion propagator invariant under the RG flow, we therefore require that 
\begin{equation}\label{eqn:zyfull}
2z_{y} = z\left(1 +\Sigma_{y}-\Sigma_{0}\right).
\end{equation}

In the bosonic propagator $\mathcal{S}_{\bm{\phi}}$ (\ref{eqn:bosonic_prop}), the coefficient
$c_y^2$ is renormalized, 
\begin{equation}\label{eqn:beta_c}
\frac{d(c_{y}^{2})}{d\ell}=c_{y}^{2}\left(z \Pi_{y}-\eta_{\phi}\right),
\end{equation}
but can be kept fixed if we chose the anomalous dimension of the bosonic order-parameter 
to be 
\begin{equation}\label{eqn:eta_phi}
\eta_{\phi}=z \Pi_{y},
\end{equation}
where the polarization shell integral is defined in Eq.~(\ref{eqn:deltacy}).

Finally, the RG equation for the Yukawa coupling $g$ at one-loop order is given by 
\begin{equation}
\label{betag}
\frac{dg}{d\ell} = g\left(\frac{z_{y}}{2}-\eta_{\psi}-\frac{1}{2}\eta_{\phi}+z\Omega \right),
\end{equation}
with $\Omega$ defined in Eq.~(\ref{eqn:Omega}).

\subsection{Fixed Point and $1/N_f$ Expansion}

\begin{figure}[t]
 \includegraphics[width=0.9\columnwidth]{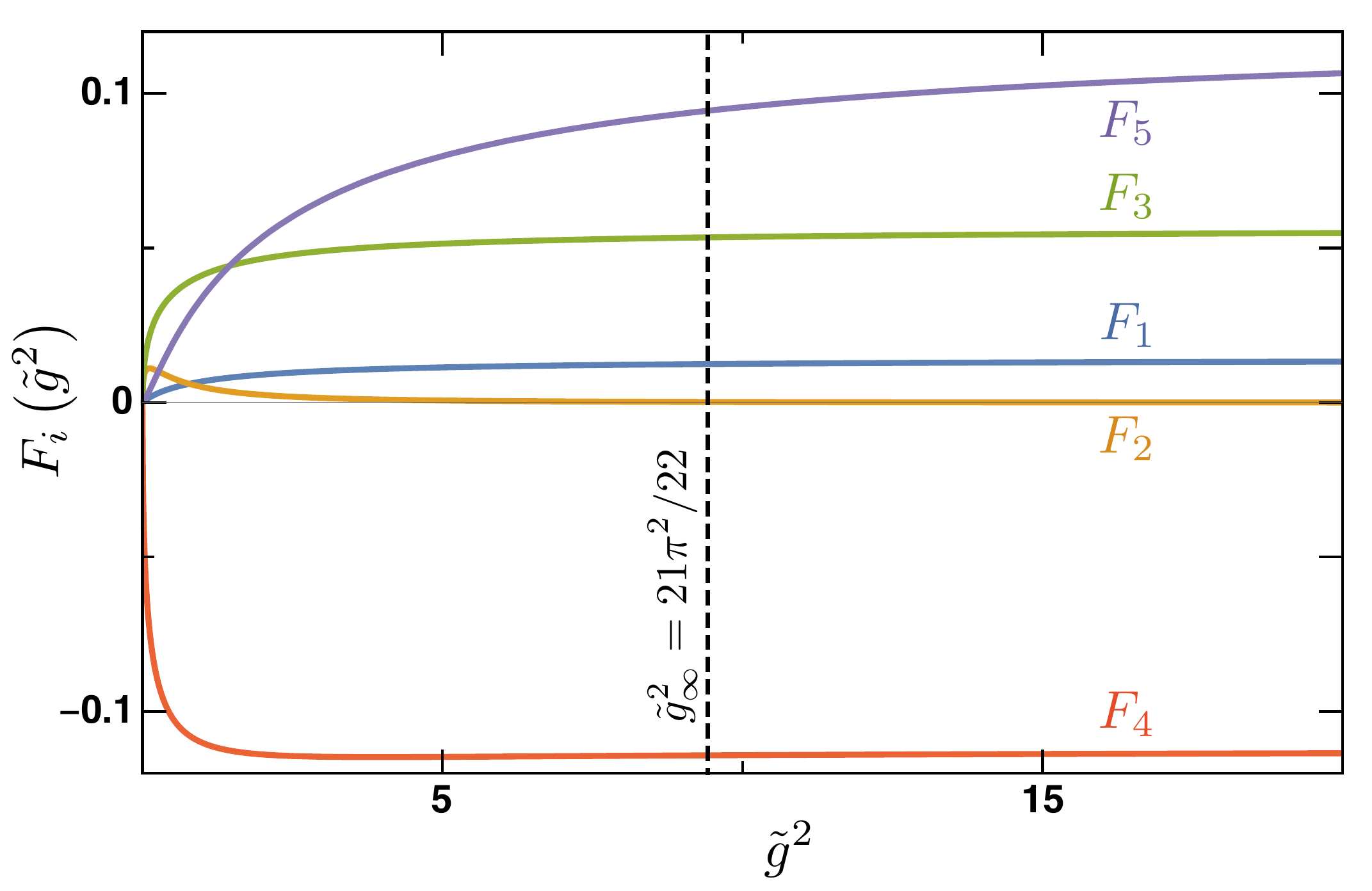}
 \caption{The integrals $F_i\left(\tilde{g}^2\right)$ as a function of the dimensionless Yukawa coupling $\tilde{g}^2 =  2N_f g^2/\sqrt{\Lambda}$. To order $1/N_f$, the critical 
 exponents at one-loop order depend on the values of $F_i$ evaluated at the critical point in the $N_f\to\infty$ limit, $\tilde{g}^2_\infty = \lim_{N_f\to\infty} \tilde{g}_*^2 = 21\pi^2/22$.}
\label{fig:f_functions}
\end{figure}

With the above choices of $\eta_{\psi}$,  $\eta_{\phi}$ and $z$ (we fix $z_y=1$ in what follows) the coefficients entering the propagators do not flow and 
can be set to 1, without loss of generality. Moreover it is convenient to define the rescaled Yukawa coupling
\begin{equation}
\tilde{g}^2 = \frac{2N_f g^2}{\sqrt{\Lambda}},
\end{equation}
as this combination enters in the argument of the functions $F_i$. These are smooth functions of $\tilde{g}^2$, as shown in Fig.~\ref{fig:f_functions}. In terms 
of the $F$ functions, the RG equation for the rescaled Yukawa coupling is given by
\begin{eqnarray}
\frac{d\tilde{g}^2}{d\ell} & = & \tilde{g}^2\left\{1 - z \left[\frac{11}{21\pi^2}  \tilde{g}^2 +\frac{N_b}{N_f }F_1\left(\tilde{g}^2\right) \right.\right.\nonumber\\
& & \left.\left. \quad\quad+\frac{2-N_b}{N_f}F_3\left(\tilde{g}^2\right)\right]\right\}
\end{eqnarray}
and from Eq.~(\ref{eqn:zyfull})
\begin{equation}
 2 = z \left\{1 - \frac{N_b}{2 N_f }\left[F_1\left(\tilde{g}^2\right) -F_2\left(\tilde{g}^2\right)\right]\right\}.
\end{equation}
The fixed point is obtained from $d\tilde{g}^2/d\ell = 0$. In the limit $N_f\to \infty$, $z=2$ and the fixed point is at
\begin{equation}
\tilde{g}^2_\infty  = \frac{21 \pi^2}{22}.
\end{equation} 

In order to obtain the leading $1/N_f$ correction to the fixed point we make an \textit{Ansatz}
\begin{equation}
\tilde{g}^2_* = \tilde{g}^2_\infty + \frac{\delta}{N_f} + \mathcal{O}(1/N_f^2).
\end{equation} 
Since all the $F$ functions have $1/N_f$ prefactors, we can replace their arguments with $\tilde{g}^2_\infty$ and define 
\begin{equation}
\alpha_i:= F_i\left( \tilde{g}^2_\infty\right).
\end{equation} 
To order $1/N_f$ we obtain 
\begin{equation}
z =   2 + \frac{N_b}{N_f} (\alpha_1-\alpha_2)
\end{equation} 
for the scaling dimension $z$ at the fixed point, and 
\begin{equation}
\frac{\tilde{g}^2_*}{\tilde{g}^2_\infty}  =   1 - \frac{N_b}{2N_f} \left(5\alpha_1-\alpha_2\right)-2 \frac{2-N_b}{N_f}\alpha_3
\end{equation} 
for the Yukawa coupling at the fixed point.
Finally, the anomalous dimensions at the critical point are
\begin{equation}\label{eqn:anomalous}
\eta_\psi =  \frac{N_b}{N_f} \alpha_1,\quad  \eta_\phi =  \frac{\tilde{g}^2_*}{\tilde{g}^2_\infty}. 
\end{equation}

The above $1/N_f$ corrections are small, and about an order of magnitude smaller than for the purely relativistic case of 
Dirac fermions in 2+1 dimensions.\cite{Janssen+14} In the latter case, the $1/N_f^2$ corrections, computed at 2-loop order,\cite{Vasiliev+93,Gracey94} are 
comparable or even larger than the $1/N_f$ ones when $N_f=1$. For semi-Dirac fermions, on the other hand, at one-loop order, the $1/N_f^2$ corrections are proportional to derivatives $F_i'\left( \tilde{g}^2_\infty \right)\simeq 10^{-4}$ and hence
 about an order of magnitude smaller than the $1/N_f$ ones. This suggests that the $1/N_f$ expansion at the physical dimension is 
 better controlled for the case of semi-Dirac fermions. However, the evaluation of two-loop diagrams would be required to investigate this further.

\subsection{Critical Exponents}
\label{sec:critical_exp} 

One can easily verify that the critical fixed point at $\tilde{g}_*$ is stable along the $\tilde{g}$ axis. Near this multicritical point there are two 
relevant perturbations, $\Delta$ and $m^2_\phi$ which are the tuning parameters for the topological phase transition and the broken symmetry 
state, respectively. 

Semi-Dirac quasiparticle excitations emerge in the semi-metallic phase when $\Delta$ is fine tuned to zero.  
Let us first consider the case that 
$\Delta$ remains equal to zero across the symmetry-breaking phase transition.  In this case the RG equation for $m_\phi^2$ is equal to 
\begin{equation}\label{eqn:m2RG}
\frac{d m^2_\phi}{d\ell}   = \left(2 - \eta_\phi\right) m^2_\phi.
\end{equation}
The correlation length is defined by the RG scale $\ell^*$ at which $m^2_\phi(\ell^*)\simeq 1$, $\xi = e^{\ell^*}$. Integrating the above differential 
equation (\ref{eqn:m2RG}),
\begin{equation}
m^2_\phi(\ell)  = m^2_\phi(0) e^{\left(2 - \eta_\phi\right) \ell}.
\end{equation}
Using that $m^2_\phi(0)\sim (g_c-g)$, we obtain $\xi \sim  (g_c-g)^{-\nu}$ with correlation length exponent $\nu  = 1/(2 - \eta_\phi)$.
From Eq.~(\ref{eqn:anomalous}), we obtain to order $1/N_f$
\begin{equation}
\nu  = 1 - \frac{N_b}{2N_f} \left(5\alpha_1-\alpha_2\right)-2 \frac{2-N_b}{N_f}\alpha_3.
\end{equation}

The electronic dispersion of semi-Dirac fermions with linear and quadratic directions is strongly anisotropic. One therefore expects that the 
order-parameter correlation inherit this anisotropy. For spatially isotropic systems, the correlations length along the imaginary time direction diverges as 
a power of the spatial correlation length, $\xi_\tau = \xi^z$, where $z$ is the dynamical exponent. With our choice $z_y=1$,  the dynamical exponent $z$ 
 sets the scaling dimension of length along the $x$ direction. We therefore have
$\xi_\tau \simeq \xi_x \simeq \xi_y^z$. The spatial anisotropy of the order-parameter correlations is therefore reflected in correlation length exponents
\begin{equation}
\label{eq.nu}
\nu_x = z\nu\quad\textrm{and}\quad \nu_y = \nu,
\end{equation}
along the $x$ and $y$ directions, respectively. In the limit $N_f\to\infty$ this gives $\nu_x=2$ and $\nu_y=1$. 

Assuming that the system satisfies hyperscaling, we can use the standard scaling relations to obtain the remaining critical exponent. The 
Josephson hyperscaling relation yields the specific heat exponent, 
\begin{eqnarray}
\alpha &  = &  2 - \nu (2z+1) \nonumber\\
& \approx &  -3 +\frac{N_b}{2 N_f} (21\alpha_1-\alpha_2) + 10\frac{2-N_b}{N_f}\alpha_3.
\end{eqnarray}
Note that the effective dimension that enters in the hyperscaling relation is equal to $D=2 z+1$ corresponding to one space-like 
and two time-like directions.
Fisher's scaling law gives the susceptibility exponent
\begin{equation}
\gamma = \left(2 - \eta_\phi\right) \nu = 1 + \mathcal{O}(1/N_f^2).
\end{equation}
We can use Rushbrooke's scaling law $\alpha+2\beta+\gamma=2$ to obtain the order-parameter critical exponent
\begin{eqnarray}
\label{eq.betaexp}
\beta & = &  1-\frac12 (\alpha+\gamma)\nonumber\\
&  \approx &  2  - \frac{N_b}{4 N_f} (21\alpha_1-\alpha_2) -5 \frac{2-N_b}{N_f}\alpha_3.
\end{eqnarray}
Finally, from the Widom identity $\gamma=\beta(\delta-1)$ we compute the field exponent
\begin{eqnarray}
\delta & = &  1+\frac{\gamma}{\beta}\nonumber\\
&  \approx &  \frac32  + \frac{N_b}{16 N_f} (21\alpha_1-\alpha_2) +\frac54 \frac{2-N_b}{N_f}\alpha_3.
\end{eqnarray}
A complete list of critical exponents with numerical values for the coefficients $\alpha_i$ can be found in 
Table \ref{table:exponents}. 

\begin{table}
\begin{tabular}[t]{|c | c |}
\hline
$z_y$   &   1 \\ \hline
$z$       &  $ 2+0.0123\;\frac{N_b}{N_f}$\\ \hline
$\eta_\psi$ & $0.0125\;\frac{N_b}{N_f}$\\ \hline
$\eta_\phi$ & $1 -0.0310\;\frac{N_b}{N_f} -0.1069\frac{2-N_b}{N_f}$\\ \hline
$\nu$ & $1 -0.0310\;\frac{N_b}{N_f} -0.1069\frac{2-N_b}{N_f}$ \\ \hline
$\alpha$ & $-3+ 0.1307\; \frac{N_b}{N_f} + 0.5345 \;\frac{2-N_b}{N_f}$ \\ \hline
$\gamma$ & $1$ \\ \hline
$\beta$ & $2 - 0.0653\; \frac{N_b}{N_f} - 0.2672 \;\frac{2-N_b}{N_f}$ \\ \hline
$\delta$ & $\frac32 + 0.0163\; \frac{N_b}{N_f} + 0.0668\;\frac{2-N_b}{N_f}$ \\ 
\hline
\end{tabular}
\caption{Critical exponents for symmetry-breaking phase transitions of semi-Dirac fermions in 2+1 dimensions, calculated at one-loop order and including $1/N_f$ corrections
in the number of fermion flavours. $N_b$ is the number of order-parameter components: $N_b=1$ for the CDW, $N_b=2$ for the superconducting and $N_b=3$ for
the SDW instabilities.}\label{table:exponents}
\end{table}

\subsection{Multicriticality}
\label{sec:multicriticality}

In Sec.~\ref{sec:critical_exp} we have summarized the universal critical behaviour of semi-Dirac fermions associated 
with spontaneous symmetry breaking due to short-range interactions. The semi-Dirac quasiparticle excitations in the 
disordered, semimetallic phase are obtained by fine tuning the system to the point of a topological phase transition between 
a Dirac semimetal with two separate relativistic Dirac points and a band insulator. In the free-fermion action (\ref{eqn:s_psi}) the 
semi-Dirac point corresponds to $\Delta=0$. Spontaneous symmetry breaking leads to the opening of a gap in the fermion 
spectrum, making it in practice challenging to ensure $\Delta=0$ 
across the transition in any real material.

Since the tuning parameters of the symmetry-breaking and topological phase transitions, $m_\phi^2$ and $\Delta$, are 
both relevant perturbations at the fixed point ($\tilde{g}=\tilde{g}_*$, $\Delta=m_\phi^2=0$),
one should view the fixed point as multicritical. The coupled RG equations for $m_\phi^2$ and $\Delta$ are 
\begin{eqnarray}
\frac{d m^2_\phi}{d\ell} & = & \left(2 - \eta_\phi\right)m^2_\phi+z\Pi_{m^2_\phi},\\
\frac{d\Delta}{d\ell} &= & \left(2 - \eta_\psi\right) \Delta+z\Sigma_\Delta
\end{eqnarray}
where $\eta_\phi$ and $\eta_\psi$ are the anomalous dimensions (\ref{eqn:anomalous}) of the fermions and bosons, 
respectively. The renormalizations $\Pi_{m^2_\phi}$ and $\Sigma_\Delta$ are coming from the bosonic self energy (Fig.~\ref{fig:b_se})
and fermionic self energy [Fig.~\ref{fig:vertex}(a)] and are equal to 
\begin{eqnarray}\label{eqn:pi_mphi}
\Pi_{m^2_\phi} & = &  \frac{2}{3\pi^2} \tilde{g}^2 \Delta \\
\label{eqn:sigma_delta}
\Sigma_\Delta  & = & \frac{N_b}{2N_f}\left[F_4\left(\tilde{g}^2\right)\Delta+ F_5\left(\tilde{g}^2\right) m^2_{\phi}\right].
\end{eqnarray}
In the last equation, we have defined the functions 
\begin{eqnarray}\label{eqn:F4}
F_{4}(x) & = & \frac{1}{\pi^{2}}\int_{0}^{\frac{\pi}{2}}d\theta\ \frac{\cos2\theta}{x^{-1}\cos\theta+\sqrt{\sin\theta}}\frac{\sin\theta}{\sqrt{\cos\theta}},\\
\label{eqn:F5}
F_{5}(x) & = & \frac{1}{\pi^{2}}\int_{0}^{\frac{\pi}{2}}d\theta\ \frac{\sqrt{\cos\theta}\sin\theta}{\big(x^{-1}\cos\theta+\sqrt{\sin\theta}\big)^2},
\end{eqnarray}
which are shown in Fig.~\ref{fig:f_functions}. The coupled RG equations are of the form 
\begin{equation}
\frac{d}{d\ell} \left(\begin{array}{c} m^2_\phi \\ \Delta \end{array}\right)= \mathbf{M}  \left(\begin{array}{c} m^2_\phi \\ \Delta \end{array}\right),
\end{equation}
where $\mathbf{M}$ is an off-diagonal matrix that we evaluate at the multicritical fixed point. The two positive eigenvalues $\theta_1$
and $\theta_2$ 
of this matrix are 
inversely proportional to correlation length exponents, $\nu_i=1/\theta_i$. Up to order $1/N_f$ we obtain
\begin{eqnarray}
\nu_1 & = & 1 - \frac{N_b}{2 N_f} \left(5\alpha_1-\alpha_2 -\frac{28}{11}\alpha_5   \right) - 2\frac{2-N_b}{N_f}\alpha_3,\\
\nu_2 & = & \frac12 + \frac{N_b}{4 N_f} \left(  \alpha_1 -  \alpha_4 -\frac{14}{11}\alpha_5\right).
\end{eqnarray}

\section{Mean-field analysis}
\label{sec.MF}

The critical exponents are expected to recover the mean-field values in the limit $N_f\to\infty$. In this
limit, the anomalous dimension of the fermion fields vanishes, $\eta_\psi=0$, indicating that Fermi-liquid
behavior is recovered. However, the anomalous dimension of the order parameter field (\ref{eqn:anomalous}) 
remains finite, $\eta_\phi=1$. This results in a correlation length exponent $\nu=1/(2-\eta_\phi)=1$ and, using standard scaling 
and hyperscaling relations, in an unusually large order-parameter exponent $\beta=2$. These exponents are 
very different from the usual mean-field ones ($\eta_\phi=0$, $\nu=\beta=\frac{1}{2}$). 
This unusual behaviour is a result of the appearance of non-analytic terms in the mean-field free energy, which lead to unconventional quantum criticality and 
 arise due to the unusual scaling of the density of states $\rho(\epsilon)\sim \sqrt{\epsilon}$ around the Fermi points.

The mean-field free energy for a gapped phase of semi-Dirac fermions is equal to 
\begin{eqnarray}\label{eq.F_MF2}
\mathcal{F}_\textrm{MF}(\phi_0) & = & \frac{1}{g}\phi_0^2 - \int_{k_x^2+k_y^4\le\Lambda^2} \frac{d^2\bm{k}}{(2\pi)^2} \sqrt{k_x^2+k_y^4+\phi_0^2},\qquad 
\end{eqnarray}
where we have rescaled $k_x$ and $k_y$ to absorb the prefactors $v$ and $1/(2m)$. Here $\Lambda$ is the UV energy cut-off. Carrying 
out the integral one obtains \cite{Uchoa2017} 
\begin{eqnarray}
\mathcal{F}_\textrm{MF}(\phi_0) & = & a (\delta g) \phi_0^2 + b |\phi_0|^{\frac{5}{2}} +\mathcal{O}(\phi_0^4).
\end{eqnarray}
with $\delta g = (g_c-g)/g_c$ and $a,b>0$. As in the case of relativistic Dirac fermions, the mean-field free
energy contains a non-analytic term, $|\phi_0|^{\frac{5}{2}}$, which arises from the evaluation of the integral in the $\bm{k}\to0$ limit. 
Minimizing $\mathcal{F}_\textrm{MF}(\phi_0)$ with respect to $\phi_0$
one obtains $|\phi_0|\sim |\delta g|^{\beta_\textrm{MF}}$ with $\beta_\textrm{MF}=2$, in agreement with the RG and scaling analysis in 
 the $N_f\to\infty$ limit, Eq.~(\ref{eq.betaexp}). In contrast, for Dirac fermions, where the density of states vanishes linearly ($\rho(\epsilon)\sim \epsilon$), the non-analytic term has the form $|\phi_0|^3$,
resulting instead in the mean-field exponent $\beta_\textrm{MF}=1$. \cite{sachdev} 

The spatial anisotropy of the system, which appears in the anisotropic dispersion of the quasiparticles, also reflects in the form of the gradient terms in the Ginzburg-Landau functional. These terms can be computed by 
allowing for small, long-wavelength modulations of the order parameter. For a finite homogeneous component $\phi_0$ one can expand 
in the momentum $\bm{q}$ of the modulation. This gives rise to terms $q_x^2 \sqrt{|\phi_0|}$ and $q_y^2 |\phi_0|^{\frac{3}{2}}$,\cite{Uchoa2017}  from which we can 
estimate the correlation lengths $\xi_x$ and $\xi_y$ along the $x$ and $y$ directions respectively. Since
\begin{equation}
\xi_x^{-2}  |\phi_0|^{\frac{1}{2}} \simeq \xi_y^{-2}  |\phi_0|^{\frac{3}{2}} \simeq  |\delta g| \phi_0^2
\end{equation}
by dimensional analysis, that leads to the quantum critical scaling 
\begin{equation} 
\label{eq.xi2x}
\xi_x ^2 \sim |\phi_0|^{-\frac{3}{2}} |\delta g|^{-1} \sim |\delta g|^{-(1+\frac{3}{2}\beta_\textrm{MF})}, 
\end{equation}
and 
\begin{equation}
\xi_y ^2 \sim |\phi_0|^{-\frac{1}{2}} |\delta g|^{-1} \sim |\delta g|^{-(1+\frac{1}{2}\beta_\textrm{MF})}.
\end{equation}
Using $\beta_\textrm{MF}=2$, this simple scaling analysis of the mean-field free energy 
recovers the correlation length exponents  $\nu_x=2$ and $\nu_y=1$ derived in Eq. (\ref{eq.nu}), in agreement with 
 the RG result in the limit $N_f\to\infty$.
 
 The anisotropic scaling of the correlation length along the $x$ and $y$ directions could have very interesting implications 
 for ordered phases in the vicinity of the quantum critical point. In general, the order parameter becomes relatively 
 softer to spatial modulations along the direction where the quasiparticles have parabolic dispersion, and more rigid in the other direction, permitting the emergence of modulated order and stripe phases. \cite{Uchoa2017} In the superconducting case,  
 the system may effectively respond to a external magnetic field as a type II superconductor in one direction and as a type I in the other. \cite{Uchoa2017} This unconventional state could stabilize stripes of magnetic flux rather than conventional vortex lattices.

\section{Discussion}
\label{sec.dis}

We have analyzed the critical behavior of quantum phase transitions in semi-Dirac fermion systems 
that are driven by short-range interactions. Here we have focussed on SDW, CDW and superconducting 
instabilities and derived the effective Yukawa actions that describe the coupling between the dynamical 
order parameter fields and the semi Dirac fermions. 

This problem was previously studied using an $\epsilon$ expansion in the number of quadratically 
dispersing directions\cite{Sur2018} and by generalizing the quadratic dispersion to $k_y^{2n}$, facilitating a $1/n$ 
expansion around the one-dimensional limit.\cite{Roy+08} 
In our complementary approach, we have avoided the tuning of dimensionality but instead introduced a large 
number $N_f$ of fermion flavors for analytic control in two spatial dimensions. Using a  one-loop 
renormalization group analysis of the effective Yukawa actions, we have computed critical exponents up to 
order $1/N_f$. 

The $1/N_f$  corrections to critical exponents depend on the peculiar form of the IR order-parameter 
propagator in 2+1 dimensions, $G_\phi^\textrm{IR} \sim\left(q_0^2+v^2 q_x^2\right)^{-\frac{1}{4}}$. This IR contribution, which is not 
renormalized by integrating out electronic UV modes, needs to be incorporated to regularize unphysical 
divergencies.\cite{Isobe2016}
Our RG equations are derived from integrating UV modes from an infinitesimal 2+1 dimensional shell in 
momentum-frequency space, followed by a rescaling of frequency and momenta to the original cut-off. 
We have treated frequency $k_0$ and the momentum $k_x$ along the relativistic direction on an equal footing
and introduced a single ``dynamical'' exponent $z$ that describes the scaling of $k_0$, $k_x$ relative 
to the quadratically dispersing direction $k_y$. One might view this as having one space-like and two time-like
directions.  

We have found that the $1/N_f$ corrections to critical exponent are smaller than for the case of Dirac fermions and seem 
to fall off more rapidly when increasing the order of $1/N_f$. This suggests that the 
$1/N_f$ expansion is well controlled even when the number of flavors $N_f$ is of the order of 1. However, calculations beyond one-loop order are required to confirm 
this conjecture. 

In the mean-field limit  $N_f\to\infty$, the anomalous dimension $\eta_\psi$ of the fermion fields vanishes,
signaling a recovery of conventional Fermi-liquid behaviour. On the other hand, the anomalous dimension of 
the order-parameter fields remains finite, $\eta_\phi=1$. This has important consequences. It gives rise to a correlation 
length exponent of $\nu=1$ instead of the conventional mean-field\ $\nu=1/2$.
Since we have defined the $y$ direction as reference length ($z_y=1,z_x=z=2$),  this corresponds to $\nu_y=z_y \nu=1$ 
and $\nu_x=z_x\nu=2$ along the two spatial directions. 

The freedom in how to define the scaling dimensions in semi-Dirac systems explains the apparent contradiction with 
Ref.~\onlinecite{Roy+08} that reports $\nu=2$. A close inspection shows that in this work the relativistic direction was 
used to define the reference length scale. In our notation this corresponds to the choice $z=z_x = 1$ and $z_y=1/2$,
leading to the same correlation length exponents  $\nu_x=2$ and $\nu_y=1$. 

The atypical correlation length exponent and the unusually large order parameter exponent $\beta = 2 +\mathcal{O}(1/N_f)$
suggest that the mean-field order-parameter theory is highly unusual. \cite{Uchoa2017} As we discussed,  the vanishing density of states
at the Fermi level gives rise to a non-analytic $|\phi_0|^{5/2}$ term in the Landau free energy. This results in $\beta_\textrm{MF}=2$,
in agreement with the RG result for $N_f\to\infty$. The highly anisotropic order-parameter 
correlations correspond to different non-analytic gradient terms in the mean-field theory.

Semi-Dirac fermions correspond to an intermediate case between Dirac fermions and ordinary metals in two spatial dimensions. The quadratic dispersion along one of the momentum directions leads to an increased density of states at low energies 
as compared to Dirac fermions, making instabilities comparatively easier due to the enlarged phase space for quantum fluctuations.  An interesting question for future studies is whether the enhanced electronic 
fluctuations near quantum critical point combined with the anisotropy of the correlations could stabilize novel phases such as modulated order.

\acknowledgements

We thank Andrew Green and Geo Jose for useful discussions. B.~U. acknowledges NSF CAREER grant 
No. DMR-1352604 for partial support. F.~K. acknowledges financial support from EPSRC under Grant EP/P013449/1.


\begin{thebibliography}{36}%
\makeatletter
\providecommand \@ifxundefined [1]{%
 \@ifx{#1\undefined}
}%
\providecommand \@ifnum [1]{%
 \ifnum #1\expandafter \@firstoftwo
 \else \expandafter \@secondoftwo
 \fi
}%
\providecommand \@ifx [1]{%
 \ifx #1\expandafter \@firstoftwo
 \else \expandafter \@secondoftwo
 \fi
}%
\providecommand \natexlab [1]{#1}%
\providecommand \enquote  [1]{``#1''}%
\providecommand \bibnamefont  [1]{#1}%
\providecommand \bibfnamefont [1]{#1}%
\providecommand \citenamefont [1]{#1}%
\providecommand \href@noop [0]{\@secondoftwo}%
\providecommand \href [0]{\begingroup \@sanitize@url \@href}%
\providecommand \@href[1]{\@@startlink{#1}\@@href}%
\providecommand \@@href[1]{\endgroup#1\@@endlink}%
\providecommand \@sanitize@url [0]{\catcode `\\12\catcode `\$12\catcode
  `\&12\catcode `\#12\catcode `\^12\catcode `\_12\catcode `\%12\relax}%
\providecommand \@@startlink[1]{}%
\providecommand \@@endlink[0]{}%
\providecommand \url  [0]{\begingroup\@sanitize@url \@url }%
\providecommand \@url [1]{\endgroup\@href {#1}{\urlprefix }}%
\providecommand \urlprefix  [0]{URL }%
\providecommand \Eprint [0]{\href }%
\providecommand \doibase [0]{http://dx.doi.org/}%
\providecommand \selectlanguage [0]{\@gobble}%
\providecommand \bibinfo  [0]{\@secondoftwo}%
\providecommand \bibfield  [0]{\@secondoftwo}%
\providecommand \translation [1]{[#1]}%
\providecommand \BibitemOpen [0]{}%
\providecommand \bibitemStop [0]{}%
\providecommand \bibitemNoStop [0]{.\EOS\space}%
\providecommand \EOS [0]{\spacefactor3000\relax}%
\providecommand \BibitemShut  [1]{\csname bibitem#1\endcsname}%
\let\auto@bib@innerbib\@empty
\bibitem [{\citenamefont {Castro~Neto}\ \emph {et~al.}(2009)\citenamefont
  {Castro~Neto}, \citenamefont {Guinea}, \citenamefont {Peres}, \citenamefont
  {Novoselov},\ and\ \citenamefont {Geim}}]{CastroNeto2009}%
  \BibitemOpen
  \bibfield  {author} {\bibinfo {author} {\bibfnamefont {A.~H.}\ \bibnamefont
  {Castro~Neto}}, \bibinfo {author} {\bibfnamefont {F.}~\bibnamefont {Guinea}},
  \bibinfo {author} {\bibfnamefont {N.~M.~R.}\ \bibnamefont {Peres}}, \bibinfo
  {author} {\bibfnamefont {K.~S.}\ \bibnamefont {Novoselov}}, \ and\ \bibinfo
  {author} {\bibfnamefont {A.~K.}\ \bibnamefont {Geim}},\ }\href {\doibase
  10.1103/RevModPhys.81.109} {\bibfield  {journal} {\bibinfo  {journal} {Rev.
  Mod. Phys.}\ }\textbf {\bibinfo {volume} {81}},\ \bibinfo {pages} {109}
  (\bibinfo {year} {2009})}\BibitemShut {NoStop}%
\bibitem [{\citenamefont {Qi}\ and\ \citenamefont {Zhang}(2011)}]{Qi2011}%
  \BibitemOpen
  \bibfield  {author} {\bibinfo {author} {\bibfnamefont {X.-L.}\ \bibnamefont
  {Qi}}\ and\ \bibinfo {author} {\bibfnamefont {S.-C.}\ \bibnamefont {Zhang}},\
  }\href {\doibase 10.1103/RevModPhys.83.1057} {\bibfield  {journal} {\bibinfo
  {journal} {Rev. Mod. Phys.}\ }\textbf {\bibinfo {volume} {83}},\ \bibinfo
  {pages} {1057} (\bibinfo {year} {2011})}\BibitemShut {NoStop}%
\bibitem [{\citenamefont {Hasan}\ and\ \citenamefont {Kane}(2010)}]{Hasan2010}%
  \BibitemOpen
  \bibfield  {author} {\bibinfo {author} {\bibfnamefont {M.~Z.}\ \bibnamefont
  {Hasan}}\ and\ \bibinfo {author} {\bibfnamefont {C.~L.}\ \bibnamefont
  {Kane}},\ }\href {\doibase 10.1103/RevModPhys.82.3045} {\bibfield  {journal}
  {\bibinfo  {journal} {Rev. Mod. Phys.}\ }\textbf {\bibinfo {volume} {82}},\
  \bibinfo {pages} {3045} (\bibinfo {year} {2010})}\BibitemShut {NoStop}%
\bibitem [{\citenamefont {Armitage}\ \emph {et~al.}(2018)\citenamefont
  {Armitage}, \citenamefont {Mele},\ and\ \citenamefont
  {Vishwanath}}]{Armitage2018}%
  \BibitemOpen
  \bibfield  {author} {\bibinfo {author} {\bibfnamefont {N.~P.}\ \bibnamefont
  {Armitage}}, \bibinfo {author} {\bibfnamefont {E.~J.}\ \bibnamefont {Mele}},
  \ and\ \bibinfo {author} {\bibfnamefont {A.}~\bibnamefont {Vishwanath}},\
  }\href {\doibase 10.1103/RevModPhys.90.015001} {\bibfield  {journal}
  {\bibinfo  {journal} {Rev. Mod. Phys.}\ }\textbf {\bibinfo {volume} {90}},\
  \bibinfo {pages} {015001} (\bibinfo {year} {2018})}\BibitemShut {NoStop}%
\bibitem [{\citenamefont {Dietl}\ \emph {et~al.}(2008)\citenamefont {Dietl},
  \citenamefont {Pi\'echon},\ and\ \citenamefont {Montambaux}}]{Dietl+08}%
  \BibitemOpen
  \bibfield  {author} {\bibinfo {author} {\bibfnamefont {P.}~\bibnamefont
  {Dietl}}, \bibinfo {author} {\bibfnamefont {F.}~\bibnamefont {Pi\'echon}}, \
  and\ \bibinfo {author} {\bibfnamefont {G.}~\bibnamefont {Montambaux}},\
  }\href {\doibase 10.1103/PhysRevLett.100.236405} {\bibfield  {journal}
  {\bibinfo  {journal} {Phys. Rev. Lett.}\ }\textbf {\bibinfo {volume} {100}},\
  \bibinfo {pages} {236405} (\bibinfo {year} {2008})}\BibitemShut {NoStop}%
\bibitem [{\citenamefont {Montambaux}\ \emph {et~al.}(2009)\citenamefont
  {Montambaux}, \citenamefont {Pi\'echon}, \citenamefont {Fuchs},\ and\
  \citenamefont {Goerbig}}]{Montambaux2009}%
  \BibitemOpen
  \bibfield  {author} {\bibinfo {author} {\bibfnamefont {G.}~\bibnamefont
  {Montambaux}}, \bibinfo {author} {\bibfnamefont {F.}~\bibnamefont
  {Pi\'echon}}, \bibinfo {author} {\bibfnamefont {J.-N.}\ \bibnamefont
  {Fuchs}}, \ and\ \bibinfo {author} {\bibfnamefont {M.~O.}\ \bibnamefont
  {Goerbig}},\ }\href {\doibase 10.1103/PhysRevB.80.153412} {\bibfield
  {journal} {\bibinfo  {journal} {Phys. Rev. B}\ }\textbf {\bibinfo {volume}
  {80}},\ \bibinfo {pages} {153412} (\bibinfo {year} {2009})}\BibitemShut
  {NoStop}%
\bibitem [{\citenamefont {Banerjee}\ \emph {et~al.}(2009)\citenamefont
  {Banerjee}, \citenamefont {Singh}, \citenamefont {Pardo},\ and\ \citenamefont
  {Pickett}}]{Banerjee2009}%
  \BibitemOpen
  \bibfield  {author} {\bibinfo {author} {\bibfnamefont {S.}~\bibnamefont
  {Banerjee}}, \bibinfo {author} {\bibfnamefont {R.~R.~P.}\ \bibnamefont
  {Singh}}, \bibinfo {author} {\bibfnamefont {V.}~\bibnamefont {Pardo}}, \ and\
  \bibinfo {author} {\bibfnamefont {W.~E.}\ \bibnamefont {Pickett}},\ }\href
  {\doibase 10.1103/PhysRevLett.103.016402} {\bibfield  {journal} {\bibinfo
  {journal} {Phys. Rev. Lett.}\ }\textbf {\bibinfo {volume} {103}},\ \bibinfo
  {pages} {016402} (\bibinfo {year} {2009})}\BibitemShut {NoStop}%
\bibitem [{\citenamefont {Rodin}\ \emph {et~al.}(2014)\citenamefont {Rodin},
  \citenamefont {Carvalho},\ and\ \citenamefont {Castro~Neto}}]{Rodin2014}%
  \BibitemOpen
  \bibfield  {author} {\bibinfo {author} {\bibfnamefont {A.~S.}\ \bibnamefont
  {Rodin}}, \bibinfo {author} {\bibfnamefont {A.}~\bibnamefont {Carvalho}}, \
  and\ \bibinfo {author} {\bibfnamefont {A.~H.}\ \bibnamefont {Castro~Neto}},\
  }\href {\doibase 10.1103/PhysRevLett.112.176801} {\bibfield  {journal}
  {\bibinfo  {journal} {Phys. Rev. Lett.}\ }\textbf {\bibinfo {volume} {112}},\
  \bibinfo {pages} {176801} (\bibinfo {year} {2014})}\BibitemShut {NoStop}%
\bibitem [{\citenamefont {Kim}\ \emph {et~al.}(2015)\citenamefont {Kim},
  \citenamefont {Baik}, \citenamefont {Ryu}, \citenamefont {Sohn},
  \citenamefont {Park}, \citenamefont {Park}, \citenamefont {Denlinger},
  \citenamefont {Yi}, \citenamefont {Choi},\ and\ \citenamefont
  {Kim}}]{Kim2015}%
  \BibitemOpen
  \bibfield  {author} {\bibinfo {author} {\bibfnamefont {J.}~\bibnamefont
  {Kim}}, \bibinfo {author} {\bibfnamefont {S.~S.}\ \bibnamefont {Baik}},
  \bibinfo {author} {\bibfnamefont {S.~H.}\ \bibnamefont {Ryu}}, \bibinfo
  {author} {\bibfnamefont {Y.}~\bibnamefont {Sohn}}, \bibinfo {author}
  {\bibfnamefont {S.}~\bibnamefont {Park}}, \bibinfo {author} {\bibfnamefont
  {B.-G.}\ \bibnamefont {Park}}, \bibinfo {author} {\bibfnamefont
  {J.}~\bibnamefont {Denlinger}}, \bibinfo {author} {\bibfnamefont
  {Y.}~\bibnamefont {Yi}}, \bibinfo {author} {\bibfnamefont {H.~J.}\
  \bibnamefont {Choi}}, \ and\ \bibinfo {author} {\bibfnamefont {K.~S.}\
  \bibnamefont {Kim}},\ }\href {\doibase 10.1126/science.aaa6486} {\bibfield
  {journal} {\bibinfo  {journal} {Science}\ }\textbf {\bibinfo {volume}
  {349}},\ \bibinfo {pages} {723} (\bibinfo {year} {2015})}\BibitemShut
  {NoStop}%
\bibitem [{\citenamefont {Kim}\ \emph {et~al.}(2017)\citenamefont {Kim},
  \citenamefont {Baik}, \citenamefont {Jung}, \citenamefont {Sohn},
  \citenamefont {Ryu}, \citenamefont {Choi}, \citenamefont {Yang},\ and\
  \citenamefont {Kim}}]{Kim2017}%
  \BibitemOpen
  \bibfield  {author} {\bibinfo {author} {\bibfnamefont {J.}~\bibnamefont
  {Kim}}, \bibinfo {author} {\bibfnamefont {S.~S.}\ \bibnamefont {Baik}},
  \bibinfo {author} {\bibfnamefont {S.~W.}\ \bibnamefont {Jung}}, \bibinfo
  {author} {\bibfnamefont {Y.}~\bibnamefont {Sohn}}, \bibinfo {author}
  {\bibfnamefont {S.~H.}\ \bibnamefont {Ryu}}, \bibinfo {author} {\bibfnamefont
  {H.~J.}\ \bibnamefont {Choi}}, \bibinfo {author} {\bibfnamefont {B.-J.}\
  \bibnamefont {Yang}}, \ and\ \bibinfo {author} {\bibfnamefont {K.~S.}\
  \bibnamefont {Kim}},\ }\href {\doibase 10.1103/PhysRevLett.119.226801}
  {\bibfield  {journal} {\bibinfo  {journal} {Phys. Rev. Lett.}\ }\textbf
  {\bibinfo {volume} {119}},\ \bibinfo {pages} {226801} (\bibinfo {year}
  {2017})}\BibitemShut {NoStop}%
\bibitem [{\citenamefont {Katayama}\ \emph {et~al.}(2006)\citenamefont
  {Katayama}, \citenamefont {Kobayashi},\ and\ \citenamefont
  {Suzumura}}]{Katayama2006}%
  \BibitemOpen
  \bibfield  {author} {\bibinfo {author} {\bibfnamefont {S.}~\bibnamefont
  {Katayama}}, \bibinfo {author} {\bibfnamefont {A.}~\bibnamefont {Kobayashi}},
  \ and\ \bibinfo {author} {\bibfnamefont {Y.}~\bibnamefont {Suzumura}},\
  }\href {\doibase 10.1143/JPSJ.75.054705} {\bibfield  {journal} {\bibinfo
  {journal} {Journal of the Physical Society of Japan}\ }\textbf {\bibinfo
  {volume} {75}},\ \bibinfo {pages} {054705} (\bibinfo {year}
  {2006})}\BibitemShut {NoStop}%
\bibitem [{\citenamefont {Pardo}\ and\ \citenamefont
  {Pickett}(2009)}]{Pardo2009}%
  \BibitemOpen
  \bibfield  {author} {\bibinfo {author} {\bibfnamefont {V.}~\bibnamefont
  {Pardo}}\ and\ \bibinfo {author} {\bibfnamefont {W.~E.}\ \bibnamefont
  {Pickett}},\ }\href {\doibase 10.1103/PhysRevLett.102.166803} {\bibfield
  {journal} {\bibinfo  {journal} {Phys. Rev. Lett.}\ }\textbf {\bibinfo
  {volume} {102}},\ \bibinfo {pages} {166803} (\bibinfo {year}
  {2009})}\BibitemShut {NoStop}%
\bibitem [{\citenamefont {Huang}\ \emph {et~al.}(2015)\citenamefont {Huang},
  \citenamefont {Liu}, \citenamefont {Zhang}, \citenamefont {Duan},\ and\
  \citenamefont {Vanderbilt}}]{Huang2015}%
  \BibitemOpen
  \bibfield  {author} {\bibinfo {author} {\bibfnamefont {H.}~\bibnamefont
  {Huang}}, \bibinfo {author} {\bibfnamefont {Z.}~\bibnamefont {Liu}}, \bibinfo
  {author} {\bibfnamefont {H.}~\bibnamefont {Zhang}}, \bibinfo {author}
  {\bibfnamefont {W.}~\bibnamefont {Duan}}, \ and\ \bibinfo {author}
  {\bibfnamefont {D.}~\bibnamefont {Vanderbilt}},\ }\href {\doibase
  10.1103/PhysRevB.92.161115} {\bibfield  {journal} {\bibinfo  {journal} {Phys.
  Rev. B}\ }\textbf {\bibinfo {volume} {92}},\ \bibinfo {pages} {161115}
  (\bibinfo {year} {2015})}\BibitemShut {NoStop}%
\bibitem [{\citenamefont {Christou}\ \emph {et~al.}(2018)\citenamefont
  {Christou}, \citenamefont {Uchoa},\ and\ \citenamefont
  {Kr\"uger}}]{Christou+18}%
  \BibitemOpen
  \bibfield  {author} {\bibinfo {author} {\bibfnamefont {E.}~\bibnamefont
  {Christou}}, \bibinfo {author} {\bibfnamefont {B.}~\bibnamefont {Uchoa}}, \
  and\ \bibinfo {author} {\bibfnamefont {F.}~\bibnamefont {Kr\"uger}},\ }\href
  {\doibase 10.1103/PhysRevB.98.161120} {\bibfield  {journal} {\bibinfo
  {journal} {Phys. Rev. B}\ }\textbf {\bibinfo {volume} {98}},\ \bibinfo
  {pages} {161120} (\bibinfo {year} {2018})}\BibitemShut {NoStop}%
\bibitem [{\citenamefont {Christou}\ \emph {et~al.}(2019)\citenamefont
  {Christou}, \citenamefont {de~Juan},\ and\ \citenamefont
  {Kr\"uger}}]{Christou+19}%
  \BibitemOpen
  \bibfield  {author} {\bibinfo {author} {\bibfnamefont {E.}~\bibnamefont
  {Christou}}, \bibinfo {author} {\bibfnamefont {F.}~\bibnamefont {de~Juan}}, \
  and\ \bibinfo {author} {\bibfnamefont {F.}~\bibnamefont {Kr\"uger}},\
  }\href@noop {} {\bibfield  {journal} {\bibinfo  {journal} {arXiv:1906.03892}\
  } (\bibinfo {year} {2019})}\BibitemShut {NoStop}%
\bibitem [{\citenamefont {Link}\ \emph {et~al.}(2018)\citenamefont {Link},
  \citenamefont {Narozhny}, \citenamefont {Kiselev},\ and\ \citenamefont
  {Schmalian}}]{Link+18}%
  \BibitemOpen
  \bibfield  {author} {\bibinfo {author} {\bibfnamefont {J.~M.}\ \bibnamefont
  {Link}}, \bibinfo {author} {\bibfnamefont {B.~N.}\ \bibnamefont {Narozhny}},
  \bibinfo {author} {\bibfnamefont {E.~I.}\ \bibnamefont {Kiselev}}, \ and\
  \bibinfo {author} {\bibfnamefont {J.}~\bibnamefont {Schmalian}},\ }\href
  {\doibase 10.1103/PhysRevLett.120.196801} {\bibfield  {journal} {\bibinfo
  {journal} {Phys. Rev. Lett.}\ }\textbf {\bibinfo {volume} {120}},\ \bibinfo
  {pages} {196801} (\bibinfo {year} {2018})}\BibitemShut {NoStop}%
\bibitem [{\citenamefont {Kotov}\ \emph {et~al.}(2012)\citenamefont {Kotov},
  \citenamefont {Uchoa}, \citenamefont {Pereira}, \citenamefont {Guinea},\ and\
  \citenamefont {Castro~Neto}}]{Kotov2012}%
  \BibitemOpen
  \bibfield  {author} {\bibinfo {author} {\bibfnamefont {V.~N.}\ \bibnamefont
  {Kotov}}, \bibinfo {author} {\bibfnamefont {B.}~\bibnamefont {Uchoa}},
  \bibinfo {author} {\bibfnamefont {V.~M.}\ \bibnamefont {Pereira}}, \bibinfo
  {author} {\bibfnamefont {F.}~\bibnamefont {Guinea}}, \ and\ \bibinfo {author}
  {\bibfnamefont {A.~H.}\ \bibnamefont {Castro~Neto}},\ }\href {\doibase
  10.1103/RevModPhys.84.1067} {\bibfield  {journal} {\bibinfo  {journal} {Rev.
  Mod. Phys.}\ }\textbf {\bibinfo {volume} {84}},\ \bibinfo {pages} {1067}
  (\bibinfo {year} {2012})}\BibitemShut {NoStop}%
\bibitem [{\citenamefont {Herbut}(2006)}]{herbutprl2006}%
  \BibitemOpen
  \bibfield  {author} {\bibinfo {author} {\bibfnamefont {I.~F.}\ \bibnamefont
  {Herbut}},\ }\href {\doibase 10.1103/PhysRevLett.97.146401} {\bibfield
  {journal} {\bibinfo  {journal} {Phys. Rev. Lett.}\ }\textbf {\bibinfo
  {volume} {97}},\ \bibinfo {pages} {146401} (\bibinfo {year}
  {2006})}\BibitemShut {NoStop}%
\bibitem [{\citenamefont {Herbut}\ \emph {et~al.}(2009)\citenamefont {Herbut},
  \citenamefont {Juri\vaccent{c}i\'{c}},\ and\ \citenamefont
  {Vafek}}]{herbutetalprb2009}%
  \BibitemOpen
  \bibfield  {author} {\bibinfo {author} {\bibfnamefont {I.~F.}\ \bibnamefont
  {Herbut}}, \bibinfo {author} {\bibfnamefont {V.}~\bibnamefont
  {Juri\vaccent{c}i\'{c}}}, \ and\ \bibinfo {author} {\bibfnamefont
  {O.}~\bibnamefont {Vafek}},\ }\href {\doibase 10.1103/PhysRevB.80.075432}
  {\bibfield  {journal} {\bibinfo  {journal} {Phys. Rev. B}\ }\textbf {\bibinfo
  {volume} {80}},\ \bibinfo {pages} {075432} (\bibinfo {year}
  {2009})}\BibitemShut {NoStop}%
\bibitem [{\citenamefont {Assaad}\ and\ \citenamefont
  {Herbut}(2013)}]{assaadherbutprx2013}%
  \BibitemOpen
  \bibfield  {author} {\bibinfo {author} {\bibfnamefont {F.~F.}\ \bibnamefont
  {Assaad}}\ and\ \bibinfo {author} {\bibfnamefont {I.~F.}\ \bibnamefont
  {Herbut}},\ }\href {\doibase 10.1103/PhysRevX.3.031010} {\bibfield  {journal}
  {\bibinfo  {journal} {Phys. Rev. X}\ }\textbf {\bibinfo {volume} {3}},\
  \bibinfo {pages} {031010} (\bibinfo {year} {2013})}\BibitemShut {NoStop}%
\bibitem [{\citenamefont {Janssen}\ and\ \citenamefont
  {Herbut}(2014)}]{Janssen+14}%
  \BibitemOpen
  \bibfield  {author} {\bibinfo {author} {\bibfnamefont {L.}~\bibnamefont
  {Janssen}}\ and\ \bibinfo {author} {\bibfnamefont {I.~F.}\ \bibnamefont
  {Herbut}},\ }\href {\doibase 10.1103/PhysRevB.89.205403} {\bibfield
  {journal} {\bibinfo  {journal} {Phys. Rev. B}\ }\textbf {\bibinfo {volume}
  {89}},\ \bibinfo {pages} {205403} (\bibinfo {year} {2014})}\BibitemShut
  {NoStop}%
\bibitem [{\citenamefont {{Gross}}\ and\ \citenamefont
  {{Neveu}}(1974)}]{grossneveuprd1974}%
  \BibitemOpen
  \bibfield  {author} {\bibinfo {author} {\bibfnamefont {D.~J.}\ \bibnamefont
  {{Gross}}}\ and\ \bibinfo {author} {\bibfnamefont {A.}~\bibnamefont
  {{Neveu}}},\ }\href {\doibase 10.1103/PhysRevD.10.3235} {\bibfield  {journal}
  {\bibinfo  {journal} {\prd}\ }\textbf {\bibinfo {volume} {10}},\ \bibinfo
  {pages} {3235} (\bibinfo {year} {1974})}\BibitemShut {NoStop}%
\bibitem [{\citenamefont {Zinn-Justin}(1991)}]{zinnjustinnpb1991}%
  \BibitemOpen
  \bibfield  {author} {\bibinfo {author} {\bibfnamefont {J.}~\bibnamefont
  {Zinn-Justin}},\ }\href {\doibase
  https://doi.org/10.1016/0550-3213(91)90043-W} {\bibfield  {journal} {\bibinfo
   {journal} {Nuclear Physics B}\ }\textbf {\bibinfo {volume} {367}},\ \bibinfo
  {pages} {105 } (\bibinfo {year} {1991})}\BibitemShut {NoStop}%
\bibitem [{\citenamefont {Cho}\ and\ \citenamefont {Moon}(2016)}]{Cho2016}%
  \BibitemOpen
  \bibfield  {author} {\bibinfo {author} {\bibfnamefont {G.~Y.}\ \bibnamefont
  {Cho}}\ and\ \bibinfo {author} {\bibfnamefont {E.-G.}\ \bibnamefont {Moon}},\
  }\href {\doibase 10.1038/srep19198} {\bibfield  {journal} {\bibinfo
  {journal} {Scientific Reports}\ }\textbf {\bibinfo {volume} {6}} (\bibinfo
  {year} {2016}),\ 10.1038/srep19198}\BibitemShut {NoStop}%
\bibitem [{\citenamefont {Isobe}\ \emph {et~al.}(2016)\citenamefont {Isobe},
  \citenamefont {Yang}, \citenamefont {Chubukov}, \citenamefont {Schmalian},\
  and\ \citenamefont {Nagaosa}}]{Isobe2016}%
  \BibitemOpen
  \bibfield  {author} {\bibinfo {author} {\bibfnamefont {H.}~\bibnamefont
  {Isobe}}, \bibinfo {author} {\bibfnamefont {B.-J.}\ \bibnamefont {Yang}},
  \bibinfo {author} {\bibfnamefont {A.}~\bibnamefont {Chubukov}}, \bibinfo
  {author} {\bibfnamefont {J.}~\bibnamefont {Schmalian}}, \ and\ \bibinfo
  {author} {\bibfnamefont {N.}~\bibnamefont {Nagaosa}},\ }\href {\doibase
  10.1103/PhysRevLett.116.076803} {\bibfield  {journal} {\bibinfo  {journal}
  {Phys. Rev. Lett.}\ }\textbf {\bibinfo {volume} {116}},\ \bibinfo {pages}
  {076803} (\bibinfo {year} {2016})}\BibitemShut {NoStop}%
\bibitem [{\citenamefont {Roy}\ and\ \citenamefont {Foster}(2018)}]{Roy+08}%
  \BibitemOpen
  \bibfield  {author} {\bibinfo {author} {\bibfnamefont {B.}~\bibnamefont
  {Roy}}\ and\ \bibinfo {author} {\bibfnamefont {M.~S.}\ \bibnamefont
  {Foster}},\ }\href {\doibase 10.1103/PhysRevX.8.011049} {\bibfield  {journal}
  {\bibinfo  {journal} {Phys. Rev. X}\ }\textbf {\bibinfo {volume} {8}},\
  \bibinfo {pages} {011049} (\bibinfo {year} {2018})}\BibitemShut {NoStop}%
\bibitem [{\citenamefont {Sur}\ and\ \citenamefont {Roy}(2018)}]{Sur2018}%
  \BibitemOpen
  \bibfield  {author} {\bibinfo {author} {\bibfnamefont {S.}~\bibnamefont
  {Sur}}\ and\ \bibinfo {author} {\bibfnamefont {B.}~\bibnamefont {Roy}},\
  }\href@noop {} {\bibfield  {journal} {\bibinfo  {journal} {arXiv:1812.05615}\
  } (\bibinfo {year} {2018})}\BibitemShut {NoStop}%
\bibitem [{\citenamefont {Uchoa}\ and\ \citenamefont {Seo}(2017)}]{Uchoa2017}%
  \BibitemOpen
  \bibfield  {author} {\bibinfo {author} {\bibfnamefont {B.}~\bibnamefont
  {Uchoa}}\ and\ \bibinfo {author} {\bibfnamefont {K.}~\bibnamefont {Seo}},\
  }\href {\doibase 10.1103/PhysRevB.96.220503} {\bibfield  {journal} {\bibinfo
  {journal} {Phys. Rev. B}\ }\textbf {\bibinfo {volume} {96}},\ \bibinfo
  {pages} {220503} (\bibinfo {year} {2017})}\BibitemShut {NoStop}%
\bibitem [{\citenamefont {JYang}\ \emph {et~al.}(2014)\citenamefont {JYang},
  \citenamefont {Moon}, \citenamefont {Isobe},\ and\ \citenamefont
  {Nagaosa}}]{Yang+14}%
  \BibitemOpen
  \bibfield  {author} {\bibinfo {author} {\bibfnamefont {B.-J.}\ \bibnamefont
  {JYang}}, \bibinfo {author} {\bibfnamefont {E.-G.}\ \bibnamefont {Moon}},
  \bibinfo {author} {\bibfnamefont {H.}~\bibnamefont {Isobe}}, \ and\ \bibinfo
  {author} {\bibfnamefont {N.}~\bibnamefont {Nagaosa}},\ }\href {\doibase
  10.1038/nphys3060} {\bibfield  {journal} {\bibinfo  {journal} {Nature
  Physics}\ }\textbf {\bibinfo {volume} {10}},\ \bibinfo {pages} {774}
  (\bibinfo {year} {2014})}\BibitemShut {NoStop}%
\bibitem [{\citenamefont {Li}\ \emph {et~al.}(2018)\citenamefont {Li},
  \citenamefont {Wang},\ and\ \citenamefont {Liu}}]{Li2018}%
  \BibitemOpen
  \bibfield  {author} {\bibinfo {author} {\bibfnamefont {X.}~\bibnamefont
  {Li}}, \bibinfo {author} {\bibfnamefont {J.-R.}\ \bibnamefont {Wang}}, \ and\
  \bibinfo {author} {\bibfnamefont {G.-Z.}\ \bibnamefont {Liu}},\ }\href
  {\doibase 10.1103/PhysRevB.97.184508} {\bibfield  {journal} {\bibinfo
  {journal} {Phys. Rev. B}\ }\textbf {\bibinfo {volume} {97}},\ \bibinfo
  {pages} {184508} (\bibinfo {year} {2018})}\BibitemShut {NoStop}%
\bibitem [{\citenamefont {Han}\ \emph {et~al.}(2018)\citenamefont {Han},
  \citenamefont {Cho},\ and\ \citenamefont {Moon}}]{Han2018}%
  \BibitemOpen
  \bibfield  {author} {\bibinfo {author} {\bibfnamefont {S.}~\bibnamefont
  {Han}}, \bibinfo {author} {\bibfnamefont {G.~Y.}\ \bibnamefont {Cho}}, \ and\
  \bibinfo {author} {\bibfnamefont {E.-G.}\ \bibnamefont {Moon}},\ }\href
  {\doibase 10.1103/PhysRevB.98.085149} {\bibfield  {journal} {\bibinfo
  {journal} {Phys. Rev. B}\ }\textbf {\bibinfo {volume} {98}},\ \bibinfo
  {pages} {085149} (\bibinfo {year} {2018})}\BibitemShut {NoStop}%
\bibitem [{\citenamefont {Han}\ \emph {et~al.}(2019)\citenamefont {Han},
  \citenamefont {Lee}, \citenamefont {Moon},\ and\ \citenamefont
  {Min}}]{Han2019}%
  \BibitemOpen
  \bibfield  {author} {\bibinfo {author} {\bibfnamefont {S.}~\bibnamefont
  {Han}}, \bibinfo {author} {\bibfnamefont {C.}~\bibnamefont {Lee}}, \bibinfo
  {author} {\bibfnamefont {E.-G.}\ \bibnamefont {Moon}}, \ and\ \bibinfo
  {author} {\bibfnamefont {H.}~\bibnamefont {Min}},\ }\href {\doibase
  10.1103/PhysRevLett.122.187601} {\bibfield  {journal} {\bibinfo  {journal}
  {Phys. Rev. Lett.}\ }\textbf {\bibinfo {volume} {122}},\ \bibinfo {pages}
  {187601} (\bibinfo {year} {2019})}\BibitemShut {NoStop}%
\bibitem [{\citenamefont {Vafek}\ and\ \citenamefont {Yang}(2010)}]{Vafek2010}%
  \BibitemOpen
  \bibfield  {author} {\bibinfo {author} {\bibfnamefont {O.}~\bibnamefont
  {Vafek}}\ and\ \bibinfo {author} {\bibfnamefont {K.}~\bibnamefont {Yang}},\
  }\href {\doibase 10.1103/PhysRevB.81.041401} {\bibfield  {journal} {\bibinfo
  {journal} {Phys. Rev. B}\ }\textbf {\bibinfo {volume} {81}},\ \bibinfo
  {pages} {041401} (\bibinfo {year} {2010})}\BibitemShut {NoStop}%
\bibitem [{\citenamefont {Vasiliev}\ \emph {et~al.}(1993)\citenamefont
  {Vasiliev}, \citenamefont {Derkachov}, \citenamefont {Kivel},\ and\
  \citenamefont {Stepanenko}}]{Vasiliev+93}%
  \BibitemOpen
  \bibfield  {author} {\bibinfo {author} {\bibfnamefont {A.}~\bibnamefont
  {Vasiliev}}, \bibinfo {author} {\bibfnamefont {S.~E.}\ \bibnamefont
  {Derkachov}}, \bibinfo {author} {\bibfnamefont {N.}~\bibnamefont {Kivel}}, \
  and\ \bibinfo {author} {\bibfnamefont {A.}~\bibnamefont {Stepanenko}},\
  }\href@noop {} {\bibfield  {journal} {\bibinfo  {journal} {Theor. Math.
  Phys.}\ }\textbf {\bibinfo {volume} {94}},\ \bibinfo {pages} {127} (\bibinfo
  {year} {1993})}\BibitemShut {NoStop}%
\bibitem [{\citenamefont {Gracey}(1994)}]{Gracey94}%
  \BibitemOpen
  \bibfield  {author} {\bibinfo {author} {\bibfnamefont {J.}~\bibnamefont
  {Gracey}},\ }\href@noop {} {\bibfield  {journal} {\bibinfo  {journal} {Int.
  J. Mod. Phys. A}\ }\textbf {\bibinfo {volume} {9}},\ \bibinfo {pages} {727}
  (\bibinfo {year} {1994})}\BibitemShut {NoStop}%
\bibitem [{\citenamefont {Sachdev}(2011)}]{sachdev}%
  \BibitemOpen
  \bibfield  {author} {\bibinfo {author} {\bibfnamefont {S.}~\bibnamefont
  {Sachdev}},\ }\href@noop {} {\emph {\bibinfo {title} {Quantum Phase
  Transitions}}},\ \bibinfo {edition} {2nd}\ ed.\ (\bibinfo  {publisher}
  {Cambridge University Press},\ \bibinfo {year} {2011})\BibitemShut {NoStop}%
\end{thebibliography}
\end{document}